\documentclass[fleqn,usenatbib]{mnras}

\usepackage[T1]{fontenc}
\usepackage{ae,aecompl}

\usepackage{graphicx}	
\usepackage{amsmath}	
\usepackage{amssymb}	
\usepackage{multicol}

\usepackage{newtxtext,newtxmath}

\newcommand{\sgBe}{B[e] supergiant}
\newcommand{\sgBeshorthand}{B[e]SG}
\newcommand{\kms}{$\rm km\,s^{-1}$}

\newcommand{\lessthanapprox}{%
  \newcommand{\p}{%
    \setbox0=\vbox{\hbox{$<$}}%
    \ht0=0.6ex \box0 }%
  \newcommand{\s}{%
    \vbox{\hbox{$\sim$}}%
  }%
  \mathrel{\raisebox{0.7ex}{%
      \mbox{$\underset{\s}{\p}$}%
    }}%
}

\title[1.583-day periodicities of GG Carinae]{GG Carinae: Discovery of orbital phase dependent 1.583-day periodicities in the B[e] supergiant binary}

\author[A. J. D. Porter et al.]{
Augustus Porter$^{1}$\thanks{E-mail: augustus.porter@physics.ox.ac.uk},
Katherine Blundell$^{1}$,
Philipp Podsiadlowski$^{1}$,
and Steven Lee$^{2,3}$
\\
$^{1}$Department of Physics, University of Oxford, Denys Wilkinson Building, Oxford, United Kingdom\\
$^{2}$Anglo-Australian Telescope, Coonabarabran NSW 2357, Australia\\
$^3$Research School of Astronomy and Astrophysics, Australian National University, Canberra, ACT 2611
}

\date{Accepted 2021 March 17. Received 2021 March 17; in original form 2020 November 23}

\pubyear{2020}

\hypersetup{draft}
\begin{document}
\label{firstpage}
\pagerange{\pageref{firstpage}--\pageref{lastpage}}
\maketitle

\begin{abstract}
GG Carinae is a binary whose primary component is a \sgBe. Using photometric data from TESS, ASAS, OMC, and ASAS-SN, and spectroscopic data from the Global Jet Watch to study visible He\,I, Fe\,II and Si\,II emission lines, we investigate the short-period variations which are exhibited in GG Car. We find a hitherto neglected periodicity of $1.583156\pm0.0002$\,days that is present in both its photometry and the radial velocities of its emission lines, alongside variability at the well-established $\sim$31-day orbital period. We find that the amplitudes of the shorter-period variations in both photometry and some of the emission lines are modulated by the orbital phase of the binary, such that the short-period variations have largest amplitudes when the binary is at periastron. There are no significant changes in the phases of the short-period variations over the orbital period. We investigate potential causes of the 1.583-day variability, and find that the observed period agrees well with the expected period of the $l=2$ f-mode of the primary given its mass and radius. We propose that the primary is periodically pulled out of hydrostatic equilibrium by the quadrupolar tidal forces when the components are near periastron in the binary's eccentric orbit ($e=0.5$) and the primary almost fills its Roche lobe. This causes an oscillation at the $l=2$ f-mode frequency which is damped as the distance between the components increases.
\end{abstract}




\section{Introduction}
\sgBe s (\sgBeshorthand s) are a class of rare stars which are not predicted by any stellar evolution models. They are characterized by hybrid spectra of hot stars with infrared excess, strong emission in Hydrogen Balmer and Helium lines, strong permitted and forbidden emission lines from a number of elements, wide absorption lines in the ultraviolet (UV) spectrum, and significant infrared excesses. These features point towards a complex circumstellar environment \citep{Zickgraf1985, Zickgraf1986, Kraus2019ASupergiants}. Currently there are only $\sim$33 confirmed \sgBeshorthand s discovered, and $\sim$25 further candidates \citep{Kraus2014DISCOVERY31, Levato2014NewCloud, Kraus2009, Krauscorrected, Kraus2019ASupergiants}. Their formation channels and the origin of the B[e] phenomenon are unclear, with some studies ascribing the phenomena to binarity \citep{Podsiadlowski2006corrected, Miroshnichenko2007, Wang2012AMBER/VLTI300} and others to non-radial pulsations \citep{Kraus2016a}. The opaque circumstellar envelopes of \sgBeshorthand s  generally preclude direct observation of photospheric absorption lines and therefore the determination of the stars' surface conditions (e.g. \citealt{Kraus2009}). The circumstellar envelopes must be formed by enhanced mass-loss or ejection, although the exact mechanism remains unknown. \sgBeshorthand s are expected to be rapid rotators \citep{Zickgraf1986}; however direct observations of the rotation speeds of \sgBeshorthand s are inconclusive \citep{Kraus2016}. \\

GG Carinae (GG Car, also known as HD 94878 and CPD-59 2855) is an enigmatic Galactic \sgBeshorthand\ binary which has been studied for over a century due to its peculiar spectroscopic and photometric properties \citep{Pickering1896HarvardSpectra., Kruytbosch1930Variability2855, Greenstein1938FourCarinae}. \cite{Lamers1998} classified GG Car as a \sgBeshorthand\ building on the work of \cite{Mcgregor1988ATOMICSTARS} and \cite{LOPES1992AStars}, noting their observation of the B[e] phenomenon in the object; its high luminosity; indications of mass-loss through P Cygni line profiles; and its hosting of a hybrid spectrum of narrow emission lines and broad absorption features. \cite{Porter2021GGPhotometry}, hereafter Paper A, using the measured parallax of GG Car in Data Release 2 from the \textit{Gaia} mission \citep{Prusti2016TheMission, Brown2018Gaia2}, refined the luminosity of the primary and used this to constrain the primary mass and radius. Table \ref{tab:ggcar_parameters} lists the primary's stellar parameters. Studies of the CO in GG Car's circumbinary disk suggest that the primary has evolved off the main sequence, but is in an early pre-red supergiant phase of its post-main sequence lifetime \citep{Kraus2009, Kraus2013, Oksala2013ProbingTransition}. However, this determination depends on the assumed rotation velocity of the primary, which is unknown.\\

\begin{table}
\centering          
\begin{tabular}{ l l}
\hline \hline
\\
    $d$ & $3.4^{+0.7}_{-0.5}$\,kpc\\
   $M_{\rm pr}$  & $24\pm4\,M_\odot$ \\ 
   $T_{\rm eff}$ & $23\,000 \pm 2000$\,K \\
   $L_{\rm pr}$ & $1.8^{+1.0}_{-0.7}\times10^5\,L_\odot$ \\
   $R_{\rm pr}$ & $27^{+9}_{-7}\,R_\odot$ \\
\end{tabular}
\caption{\textit{Gaia} DR2 distance, $d$, and stellar parameters of the primary in GG Car, where $M_{\rm pr}$ is the mass of the primary, $T_{\rm eff}$ is the effective temperature of the primary, $L_{\rm pr}$ is the luminosity of the primary, and $R_{\rm pr}$ is the radius of the primary. All values taken from Paper A \protect\citep{Porter2021GGPhotometry} except $T_{\rm eff}$, which is taken from \protect\cite{Marchiano2012}.} 
\label{tab:ggcar_parameters}      
\end{table}

Paper A investigates the variability of GG Car over its orbital period in photometry and Global Jet Watch (GJW) spectroscopy. We found that the photometric variations are continuous over the orbital period with one maximum and one minimum, and also that the He\,I, Fe\,II and Si\,II emission lines in GG Car's visible spectrum originate in the wind of the \sgBeshorthand\ primary. We then determined an accurate orbital solution of the binary in GG Car, and found the orbit is significantly eccentric ($e=0.50$).  Paper A shows that the system is brightest in the $V$-band at periastron, and that the photometric variations of the system at the orbital period may be described by enhanced mass transfer at periastron, with the secondary accreting the wind of the primary. The full orbital solution is given in Table \ref{tab:ggcar_porter_orbital_parameters}. Orbital phases in this study are calculated
\begin{equation}
\label{eq:orbital_phase}
    \text{Orbital phase} = \frac{T - T_{\rm peri}}{P},
\end{equation}
\noindent where $T$ is time in JD, $T_{\rm peri}$ is time of periastron passage, and $P$ is orbital period.\\

\begin{table}
\centering          
\begin{tabular}{ l l}
\hline \hline
\\
   $P$ & $31.01^{+0.01}_{-0.01}$\, days \\
   $K$ & 	$48.57^{+2.04}_{-1.87}$\,\kms\\
   $\omega$ &$339.87^{+3.10}_{-3.06}$\,$^\circ$\\
   $e$ & $0.50^{+0.03}_{-0.03}$ \\
   $T_{\rm peri}$ & JD\,$2452069.36\pm1.30$ \\ \\
   
   $M_{\rm sec}$ & $7.2^{+3.0}_{-1.3}\,M_\odot$ \\
   $a$ & $0.61\pm0.03$\,AU\\
   
\end{tabular}
\caption{Orbital parameters of the \sgBeshorthand\ primary in GG Car found by Paper A \citep{Porter2021GGPhotometry}. $P$ is the orbital period, $K$ is the amplitude of the radial velocity, $\omega$ is the argument of periastron, $e$ the orbital eccentricity, $T_{\rm peri}$ is the time of periastron. $M_{\rm sec}$ is the inferred mass of the secondary, and $a$ is the resulting orbital separation. } 
\label{tab:ggcar_porter_orbital_parameters}      
\end{table}

Early time-series photometry studies noticed that GG Car displays significant intra-night variability, separate from its variability over its $\sim$31 day orbital period \citep{Kruytbosch1930Variability2855, Greenstein1938FourCarinae}. \cite{Gosset1984}, through Fourier analysis, found an indication of a periodicity at $\sim$1.6\,days in the system's photometry. This led them to state that one of the GG Car components is a variable, but no further analysis was undertaken and this periodicity has been neglected since that publication. \cite{Krtickova2018AnEnvelopes} were unable to determine a clear UV lightcurve of GG Car over the orbital period, presumably due to variability in the system; they conclude that the binary component that is brightest in the UV is the variable, but do not find a period. \\

In this study, we investigate this short-period variability of GG Car in detail, in both photometry and spectroscopy. The structure of this paper is as follows: Section \ref{sec:observations} introduces the $V$-band and TESS photometry, and the Global Jet Watch spectroscopy of GG Car; Section \ref{sec:variability} studies the variability of the system's photometry and emission lines' radial velocities; Section \ref{sec:orbital_phase_dependence_of_short_variability} investigates the relationship between the amplitude of the short-period variability in the system and the orbital phase of the binary; Section \ref{sec:discussion} presents our discussions; and Section \ref{sec:conclusions} presents our conclusions.

\section{Observations}
\label{sec:observations}
\subsection{$V$-band photometric observations}
\label{sec:v_band_observations}

$V$-band photometric data of GG Car are available from the All Sky Automated Survey (ASAS, \citealt{Pojmanski2002TheHemisphere, Pojmanski2004TheSurvey}), the Optical Monitoring Camera (OMC) aboard the INTEGRAL satellite \citep{Mas-Hesse2003OMC:INTEGRAL}, and the All Sky Automated Survey for Supernovae (ASAS-SN, \citealt{Shappee2014THE2617, Kochanek2017TheV1.0}). Each of these surveys uses standard Johnson V-filters, centred at 550\,nm and with a full width half maximum of 88\,nm. Further details of the $V$-band observations used in this study for each survey are given in Paper A.\\

\subsection{TESS photometry}
The Transiting Exoplanet Survey Satellite (TESS, \citealt{Ricker2014TransitingSatellite}) is a mission geared towards discovering new exoplanet candidates; however, its high-cadence and high-precision photometry of the majority of the sky means that it has proved a valuable resource for stellar astrophysics. The satellite is in a highly-elliptical 13.7-day orbit around Earth, and observes the sky in 26 partially overlapping ``Sectors'', each Sector being observed for roughly one month. The passband filter has an effective wavelength of 7\,500\,\AA\ and a width of 4\,000\,\AA; this wide bandpass is roughly centred on the Johnson $I_c$ band, but also encompasses the $R_c$ and $z$ bands. The filter therefore transmits to longer wavelengths than the $V$-band surveys described in Section \ref{sec:v_band_observations}. TESS is able to create exquisite light curves for objects whose Johnson V-magnitude lies between 3--12\,mags. \\

400\,000 pre-selected sources have had reduced photometric data at two minute cadence released, of which GG Car is unfortunately not a member. However, unreduced full-frame image (FFI) data with a sampling rate of 30 minutes are available for any source which lies within one of TESS's sectors. GG Car is located in TESS Sectors 10 and 11 which were observed from 2019-03-26 to 2019-05-21, covering almost two full orbital cycles of the binary, and its mean $V$-band magnitude of $\sim$8.6\,mag places it ideally within the observing limits of TESS. We reduced the TESS FFI data using the \texttt{eleanor} framework \citep{Feinstein2019Eleanor:Images}. \\

The pixel scale of TESS is 21 arcseconds per pixel, with a point-spread-function of a similar scale. This presents a problem for GG Car since it is only separated by 49\,arcseconds from its nearest neighbour, V413 Car. \texttt{eleanor} is able to minimise the impact that this may have by choosing optimal apertures and PSF modelling. We reduce the FFI data to 15$\times$15 pixel ``postage stamps'', and model the PSFs of both GG Car and V413 Car as Moffat profiles. We block the brightest 20\% of pixels away from the target, which effectively masks background stars, aiding the background subtraction. The data are converted to TESS magnitudes, using the mean magnitude of 7.696 taken from the TESS Input Catalogue. The uncertainties of the individual data points of the TESS data are very low, of the order $10^{-4}$\,mag.\\

To ensure that the features seen in the TESS light curve of GG Car are real and not instrumental artefacts, we extracted the data for three similarly bright stars which were observed nearby on the same CCD as GG Car (V413 Car, AG Car, and HD\,94961) using similar methods. The light curves of these other stars do not display the same features as those observed in GG Car.\\

\begin{figure} 
  \centering
    \includegraphics[width=0.5\textwidth]{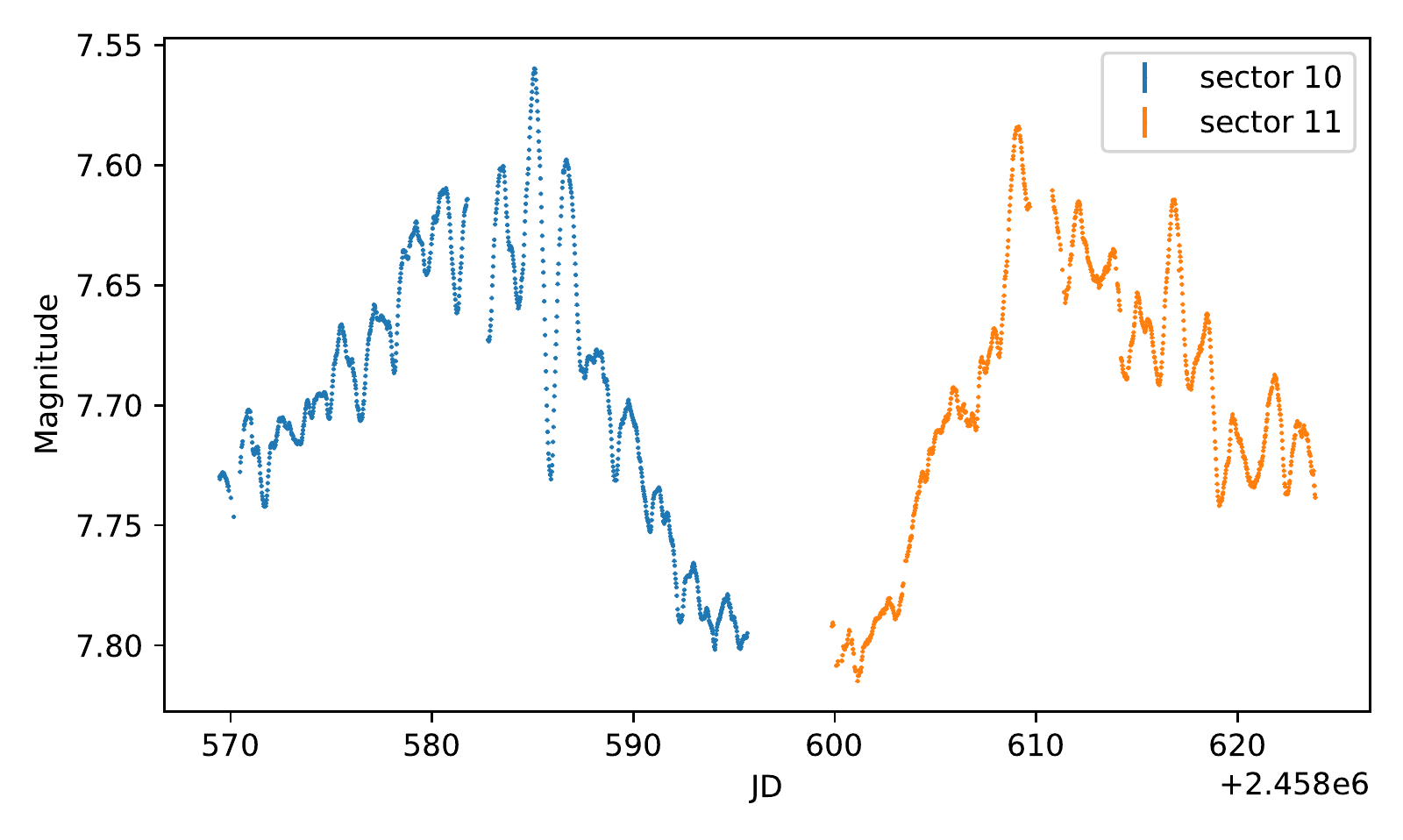}
    \caption{TESS lightcurve of GG Car.}
    \label{fig:ggcar_tess}
\end{figure}

\subsection{Global Jet Watch spectroscopy}
The Global Jet Watch (GJW) has been collecting mid-resolution (R$\sim$4\,000) optical spectroscopic data on a variety of objects, including GG Car which it has been observing since early 2015. GJW is an array of five telescopes separated in longitude which take optical spectra from $\sim$\,5\,800\,-\,8\,400\,\AA. Our observations of GG Car have exposure times of either 1000 or 3000 seconds. This is due to the dominant brightness of H-alpha. H-alpha is saturated in the 1000 and 3000\,s exposures. The spectra are barycentric corrected using heliocentric velocities calculated with the \texttt{barycorrpy} package \citep{Kanodia2018Pythonbarycorrpy}. In this study, all spectra are normalised by the local continuum. The GJW spectra studied are further described in Paper A. \\

\section{The short-period variability of GG Car}
\label{sec:variability}

\subsection{Photometric variability}
\label{sec:photometric_variability}
Figure \ref{fig:ggcar_tess} displays the TESS lightcurve of GG Car. The TESS data cover nearly two orbital cycles of the binary of GG Car with high precision and cadence. Both the longer-term variation at the orbital period and the shorter period variations along the lightcurve are apparent. It is also clear that the amplitude of the short period variation changes over the epoch of observation, with the amplitudes being correlated with the brightness of the system.\\

\begin{figure} 
  \centering
    \includegraphics[width=0.5\textwidth]{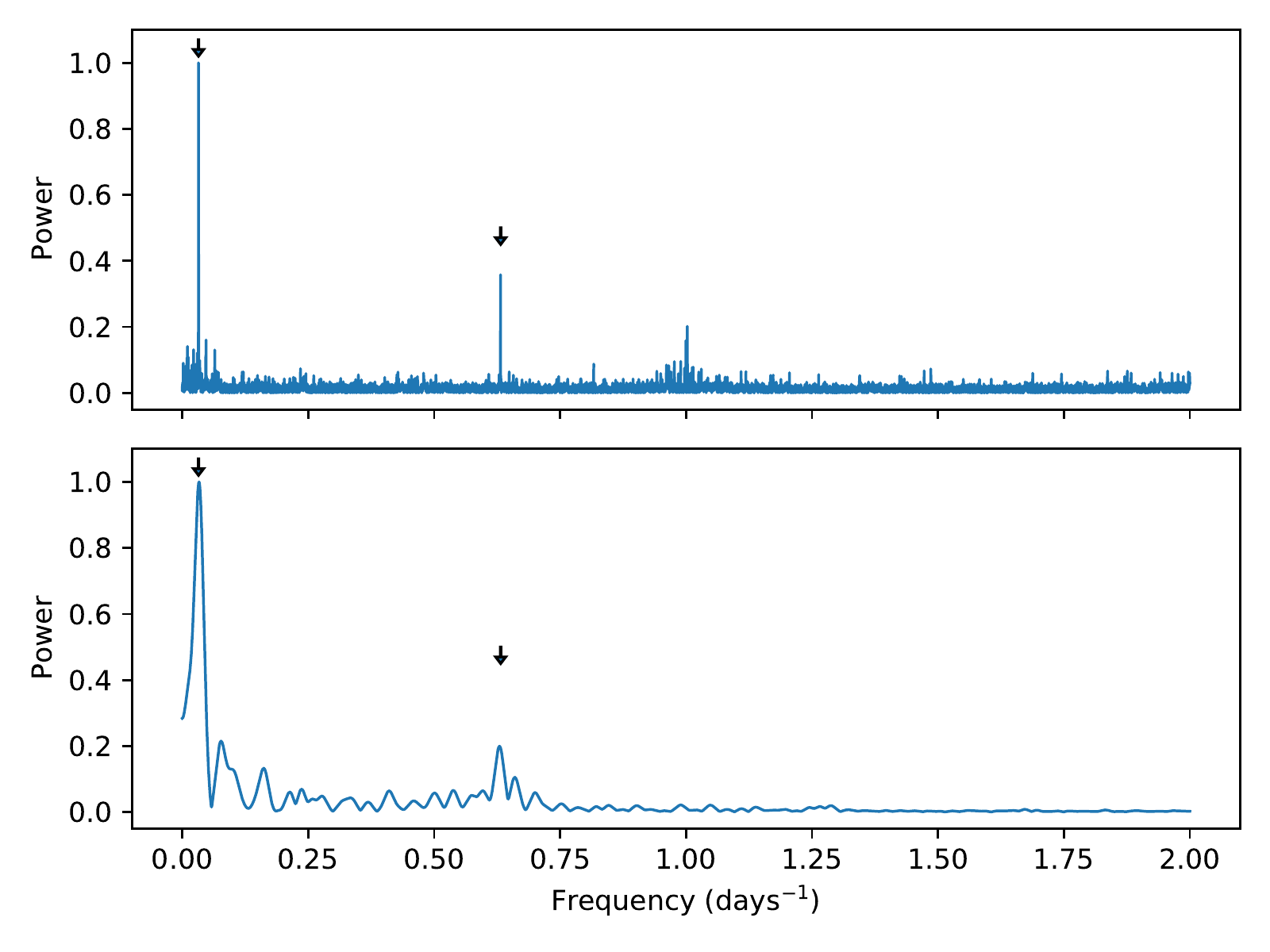}
    \caption{Fourier power spectra of the $V$-band photometry of GG Car (top panel) and the TESS photometric data (bottom panel). The frequencies corresponding to the orbital period and the short period are denoted by small arrows and presented in Table \ref{tab:fourier_period_estimates}. The periodograms are normalised by the peak power.}
    \label{fig:gg_car_v_band_photometry_power}
\end{figure}

Figure \ref{fig:gg_car_v_band_photometry_power} displays the Fourier power spectrum of the ASAS, ASAS-SN, and OMC $V$-band photometry of GG Car in the top panel, and the power spectrum of the TESS photometry in the bottom panel. These power spectra, along with all subsequent power spectra in this study, were calculated using the CLEAN algorithm of \cite{Roberts1987}, which deconvolves the Fourier transform of the data from the Fourier transform of the observational aperture function, thereby overcoming the artefacts that inevitably arise from transforming irregularly sampled data. \\

The photometric periodograms are dominated by the $\sim$31-day orbital period of the binary, however there are other periods present in the power spectrum. Most notable is the significant peak at a higher frequency of $\sim$0.632 days$^{-1}$, corresponding to a $\sim$1.583-day period. This is the same higher-frequency period noted in \cite{Gosset1984}. The $V$-band data were all taken from ground-based surveys, with the exception of OMC, so there is a peak complex in this power spectrum around 1\,days$^{-1}$ due to aliasing effects arising from the Earth's rotation period. There is no such complex in the TESS periodogram, since it is a space-based mission.\\

\begin{table}
\centering          
\begin{tabular}{p{0.15\textwidth} p{0.15\textwidth}}
Data source & Periodicities \\
\hline
\\
   $V$-band photometry & \begin{tabular}{@{}l@{}}
$31.028\pm0.07$\,days\\
$1.583156\pm0.0002$\,days\\
   \end{tabular} \\ \\
   
   TESS photometry & \begin{tabular}{@{}l@{}}
$30.2\pm6.3$\,days\\
$1.588\pm0.025$\,days\\
   \end{tabular} \\ \\
\end{tabular}
\caption{The periodicities found in the photometric data of GG Car.} 
\label{tab:fourier_period_estimates}  
\end{table}

We calculate the frequency and estimate the uncertainty of the peaks in the power spectra by fitting Gaussians to them, and thereby utilising the centroids and the standard deviations of the peaks as the frequencies and their uncertainties, respectively. Table \ref{tab:fourier_period_estimates} presents the periodicities found from the photometric data. The $\sim$31-day orbital period derived from the photometry is consistent with the spectroscopic period presented in Paper A within their respective uncertainties. Therefore for the rest of this paper we will calculate orbital phases using the spectroscopic period and ephemeris of Paper A, which is more precise at $31.01\pm0.01$\,days, to match the orbital phase of the binary. The values of the short period used in this study is the 1.583156$\pm$0.0002\,day value from the $V$-band photometry, as it is more precise than the TESS determination, and we use this period for both photometry and spectroscopy. We therefore calculate the phases of the short period by
\begin{equation}
\label{eq:short_period_phase}
    \text{phase} = \frac{T - T_{\rm peri}}{1.583156},
\end{equation}
\noindent where $T$ is the JD of observation. We calculate the phases relative to $T_{\rm perid}$, though the choice of this reference time is arbitrary for the short period and without physical significance. \\

\begin{figure} 
  \centering
    \includegraphics[width=0.5\textwidth]{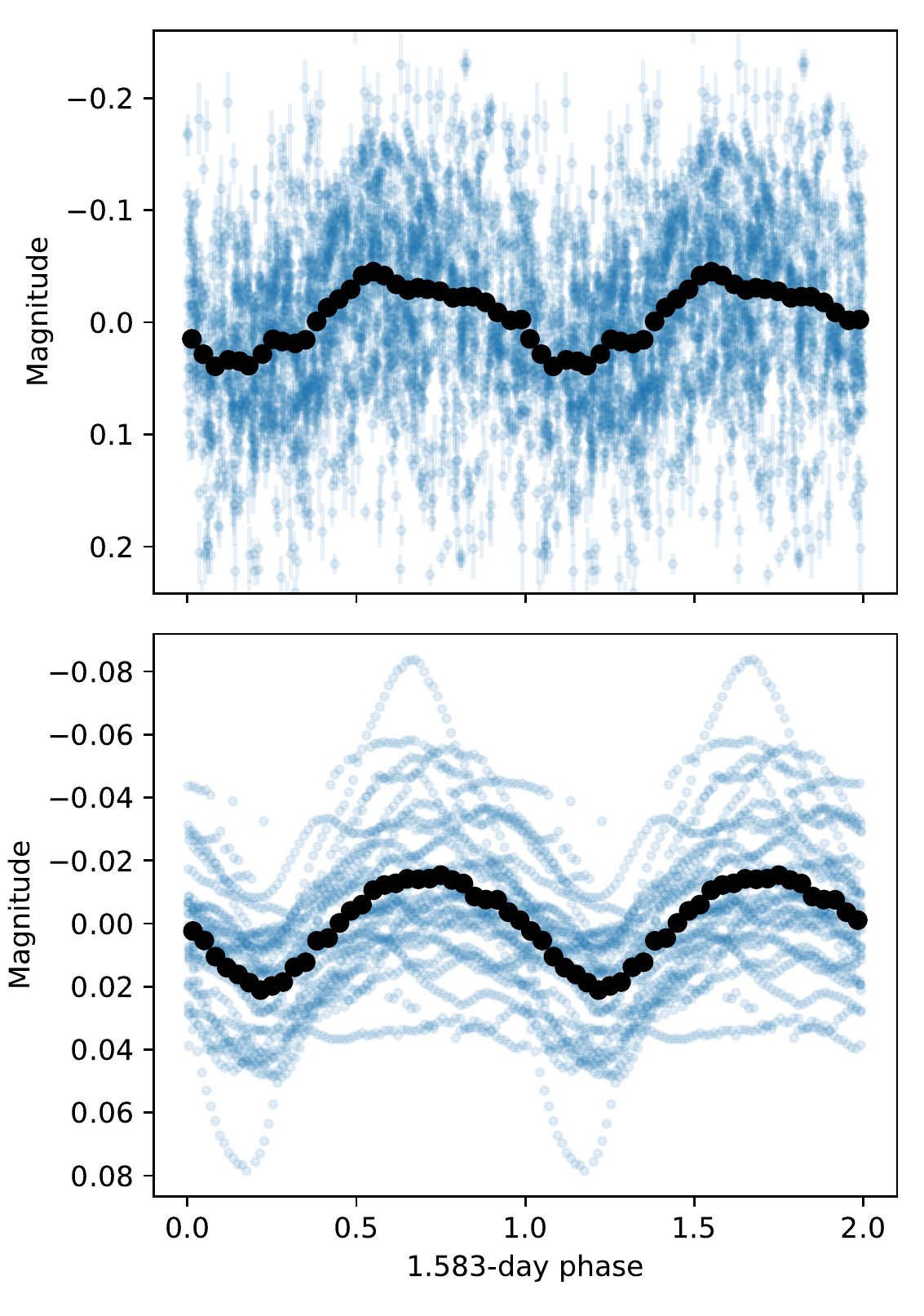}
    \caption{Top panel: $V$-band photometry of GG Car, folded by the short 1.583-day period using Equation \ref{eq:short_period_phase}. The data have had variations at the 31-day orbital period subtracted before folding. Black points indicate the average value in 30 bins by phase. Bottom panel: same as the top panel, except for the TESS photometric data, and the black points denote the average in 35 bins by phase.}
    \label{fig:v_band_tess_folded_combined}
\end{figure}

Figure \ref{fig:v_band_tess_folded_combined} displays the $V$-band and TESS photometry data folded according to Equation \ref{eq:short_period_phase}, once the variations at the orbital period are subtracted. Black points average values by phase bin. The short period is very clear to see in both the $V$-band and the TESS data and the averages show that the variations agree in phase indicating an accurate and persistent period, as the two datasets cover nearly 19 years of observation. There is, however, considerable scatter in the folded data. A cause of this scatter is, as discussed in Paper A, because the photometric variations for each 31-day orbital cycle are not identical, but they vary in shape and depth. This can also be seen in the TESS photometric data in Figure \ref{fig:ggcar_tess}; the middle minimum is deeper than the minima at the start and end of the observational period. Another cause of the scatter is the variable amplitude of the 1.583-day variations, which is very clear in the TESS data. This amplitude modulation is further discussed in Section \ref{sec:orbital_phase_dependence_of_short_variability}.\\

\subsection{Spectroscopic variability}
We now turn to spectroscopic variability, as observed by the Global Jet Watch. The variability in this section focuses on the radial velocities (RVs) of emission lines in the visible spectrum of GG Car. We use the same spectra and Gaussian fitting methods to extract the RVs of the emission lines as Paper A, and we refer the reader to that study for details of the methodology. Paper A has shown that Gaussian fitting is a robust method to extract emission lines centers, amongst other studies (e.g. \citealt{Blundell2007FluctuationsTimescales, Grant2020Uncovering140}). The 1.583-day periodicity is detected in the RVs of the He\,I emission lines, and in some Si\,II and Fe\,II emission lines.\\

\begin{figure} 
  \centering
    \includegraphics[width=0.5\textwidth]{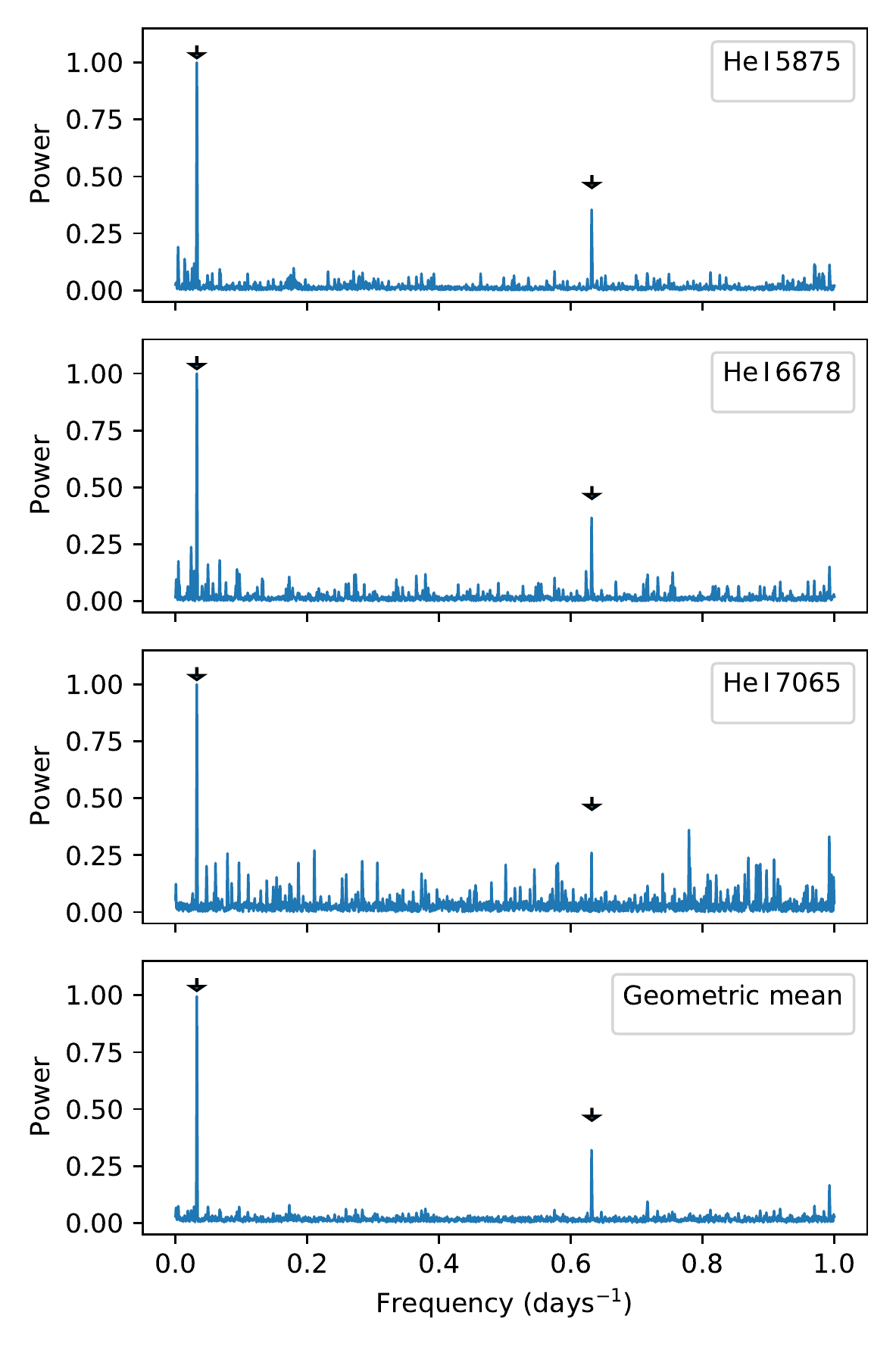}
    \caption{Periodograms of the radial velocities of the emission components of the He\,I lines. The bottom panel displays the geometric mean of the three periodograms. Arrows indicate the frequencies of the orbital period and the new 1.583-day short period. The periodograms are normalised by the peak power, which for these lines are all at the orbital period.}
    \label{fig:helium_lines_periodograms}
\end{figure}

Figure \ref{fig:helium_lines_periodograms} displays the Fourier power spectra for the RVs of the emission of the three visible He\,I emission lines. The bottom panel displays the geometric mean of the three power spectra to detect common periodicities across the line species. Taking the geometric mean leads to spurious peaks in the individual periodograms being suppressed, and common periodicities which exist across all line species to be promoted. This allows us to observe whether real periodicities exist in noisy periodograms. As with the photometric power spectra in Figure \ref{fig:gg_car_v_band_photometry_power}, the He\,I variability is dominated by the orbital period, and each of the lines also has a peak at the short-period 1.583-day frequency. There are no common significant peaks at other periods, as shown in the geometric mean of the power spectra.\\

\begin{figure} 
  \centering
    \includegraphics[width=0.5\textwidth]{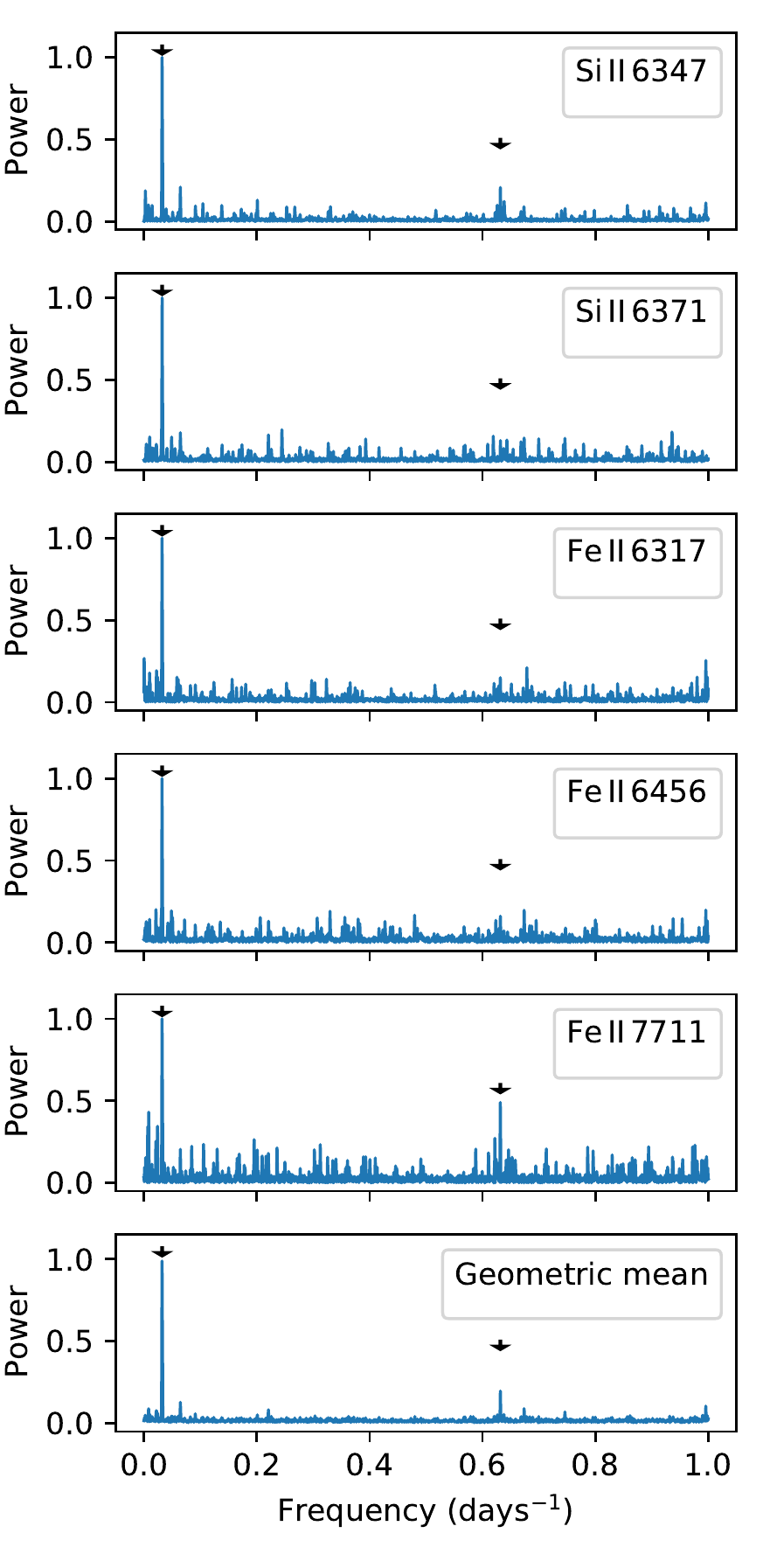}
    \caption{Same as Figure \ref{fig:helium_lines_periodograms}, except for the Si\,II and Fe\,II emission lines which display the 1.583-day variation.}
    \label{fig:metal_lines_short_period_periodograms}
\end{figure}

Periodograms were similarly calculated for the RVs extracted from Fe\,II and Si\,II lines. The majority of these lines are also dominated by variations at the orbital period, and a minority show indications of the 1.583-day variations. The metal lines which show an indication of RV variability at the short period frequency are the Fe\,II 6317.4, 6456.4, 7712.4 and Si\,II 6347.1, 6371.4 lines, and their periodograms are displayed in Figure \ref{fig:metal_lines_short_period_periodograms}. Even though some of the individual periodograms in Figure \ref{fig:metal_lines_short_period_periodograms} do not have a significant peak at the short-period compared to the noise of the periodogram, a significant peak survives at this frequency in the geometric mean of the power spectra. Lines studied in Paper A which show no indication of variability at 1.583 days in their power spectra are H-alpha, Fe\,II\,5991.3711, 6383.7302, 6432.676, 6491.663, 7513.1762, and Si\,II\,5957.56, 5978.93. Their periodograms are not included in this paper as they do not enlighten, and these lines without the short-period variability are not investigated further in this study.\\

\begin{figure} 
  \centering
    \includegraphics[width=0.5\textwidth]{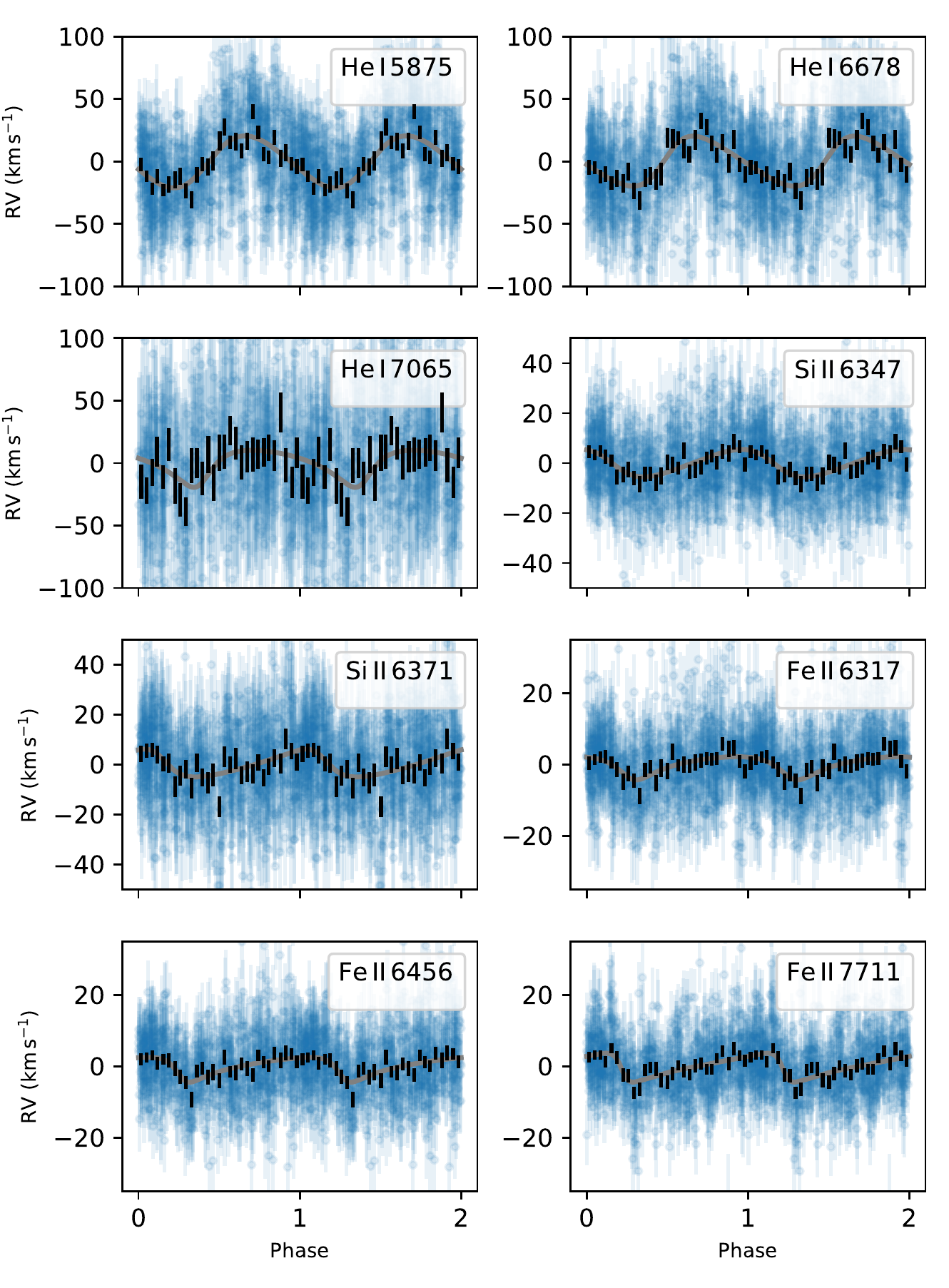}
    \caption{Radial velocity variations of the emission lines at the 1.583-day short period. Phases are calculated using Equation \ref{eq:short_period_phase}. The variations at the 31-day orbital period have been subtracted from the data of each line, to remove scatter. The data are split into 30 phase bins for each line species and black error bars indicate the weighted mean and standard error of the weighted mean for the radial velocities in each bin. The error bars of the data are both the uncertainties from the Gaussian fitting routine and the jitter, added in quadrature.}
    \label{fig:he_lines_rv_fits}
\end{figure}

Figure \ref{fig:he_lines_rv_fits} displays the RV variations of the emission lines which display variability at the 1.583-day period. The data are binned into 30 bins by phase, and the weighted mean and standard error on the weighted mean in each bin are given by the black error bars. The RV variations have different amplitudes and profiles for each line species, similar to what was found in Paper A with the RV variations at the orbital period. It is unclear why certain lines display the short-period variations whilst others do not, even lines which arise from the same ionisation species. For the He\,I lines, which display the short-period variations, comparing their RV curves to the photometric variations the phase of the maximal blueshift of the 1.583-day variations roughly corresponds with the phase of minimal brightness at this period.\\

\begin{figure} 
  \centering
    \includegraphics[width=0.5\textwidth]{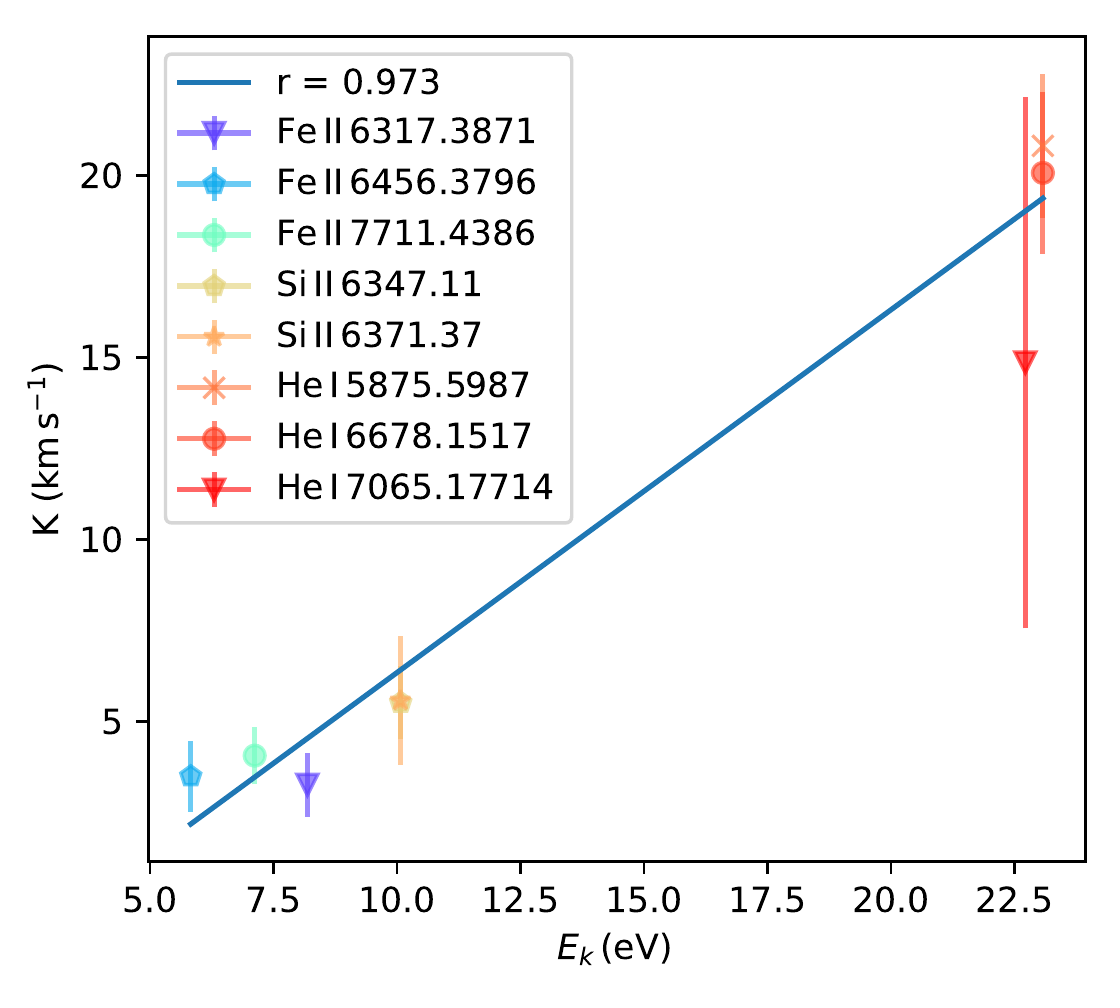}
    \caption{Amplitude, $K$, of the RV variations of the emission lines at the 1.583-day period against $E_k$, the energy of the initial excited state leading to the lines. The correlation coefficient, $r$, weighted by the inverse of the square of the error, is quoted in the legend.}
    \label{fig:k_v_ek_1583}
\end{figure}

Similar to the findings in Paper A at the orbital period, the amplitude of the short-period RV variations varies according to line species, with He\,I having the largest variations and Fe\,II the smallest. Figure \ref{fig:k_v_ek_1583} plots the amplitude, $K$, of the RV data against the energy of the upper atomic state of the transition, $E_k$. A clear correlation of $K$ and $E_k$ is evident, even with the small number of lines available. The method of determining $K$ for the RV variations in described in Appendix \ref{sec:appendix_k}. \\

\subsection{Orbital-phase dependence of short-period variability} \label{sec:orbital_phase_dependence_of_short_variability}

\begin{figure} 
  \centering
    \includegraphics[width=0.5\textwidth]{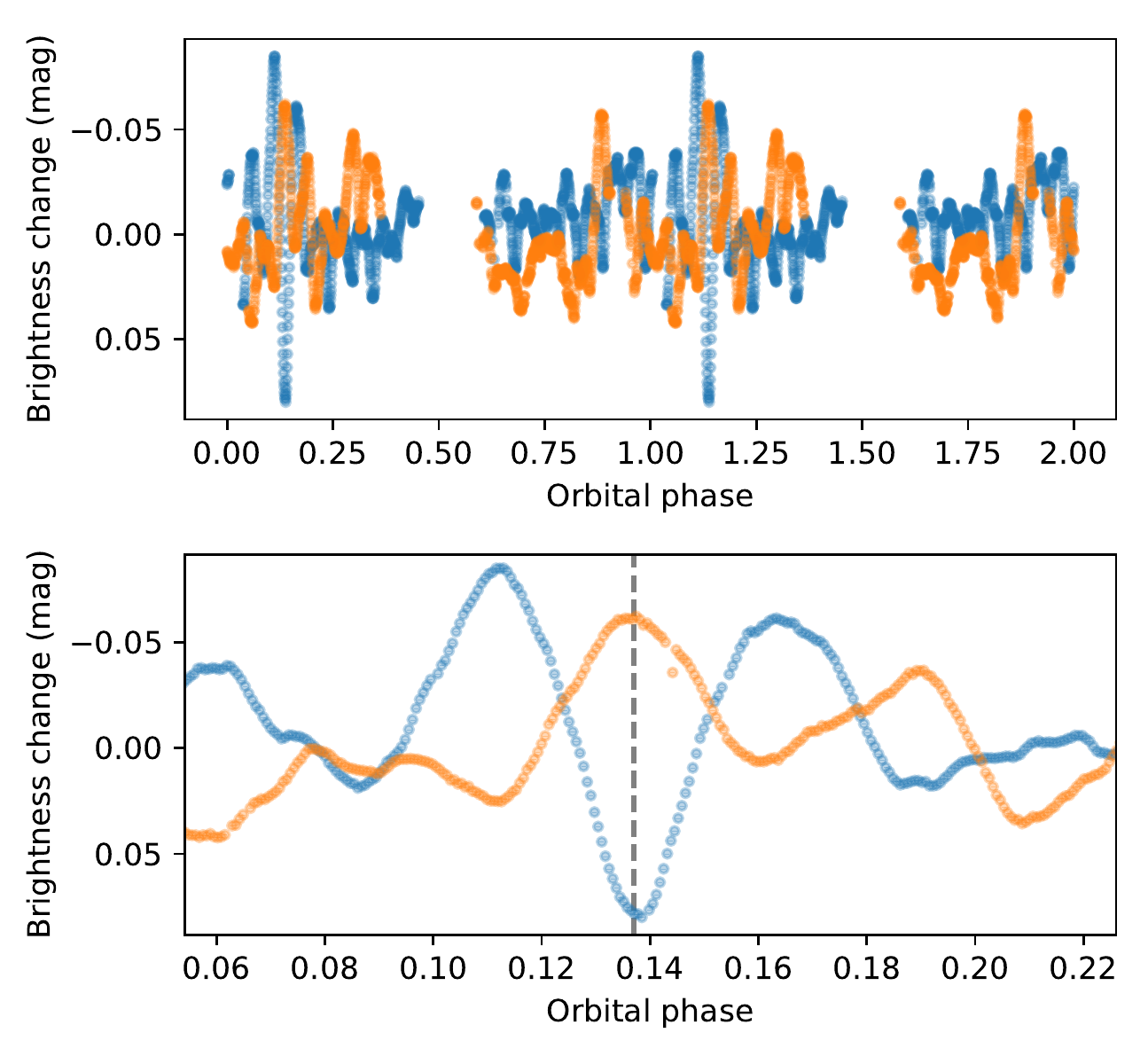}
    \caption{TESS photometric data of GG Car with the orbital period variations subtracted, leaving the short-period variations, then folded over the orbital period using Equation \ref{eq:orbital_phase}. Sector 10 data are in blue, and sector 11 in orange. The bottom panel zooms in on the largest variations around phase 0.137, where a dashed vertical line is drawn.}
    \label{fig:tess_orbital_subtracted}
\end{figure}

Here we focus on how the amplitudes of the short-period variability are modulated by the phase of the binary orbit. Figure \ref{fig:ggcar_tess} shows that, in the TESS observations of GG Car, both the orbital and 1.583-day photometric variations are not uniform, but they vary in both amplitude and profile. Figure \ref{fig:tess_orbital_subtracted} shows the TESS photometric data once they have had the orbital period variations subtracted, leaving only the short-period variability, and then folded by orbital phase using the orbital solution given in Table \ref{tab:ggcar_porter_orbital_parameters}. The largest variations in both observed orbital cycles occurs at around phase 0.12, with those variations being zoomed in the bottom panel. The variations are smallest around phase 0.6, though there is a data gap between phase 1.45--1.55. \\

\begin{figure} 
  \centering
    \includegraphics[width=0.5\textwidth]{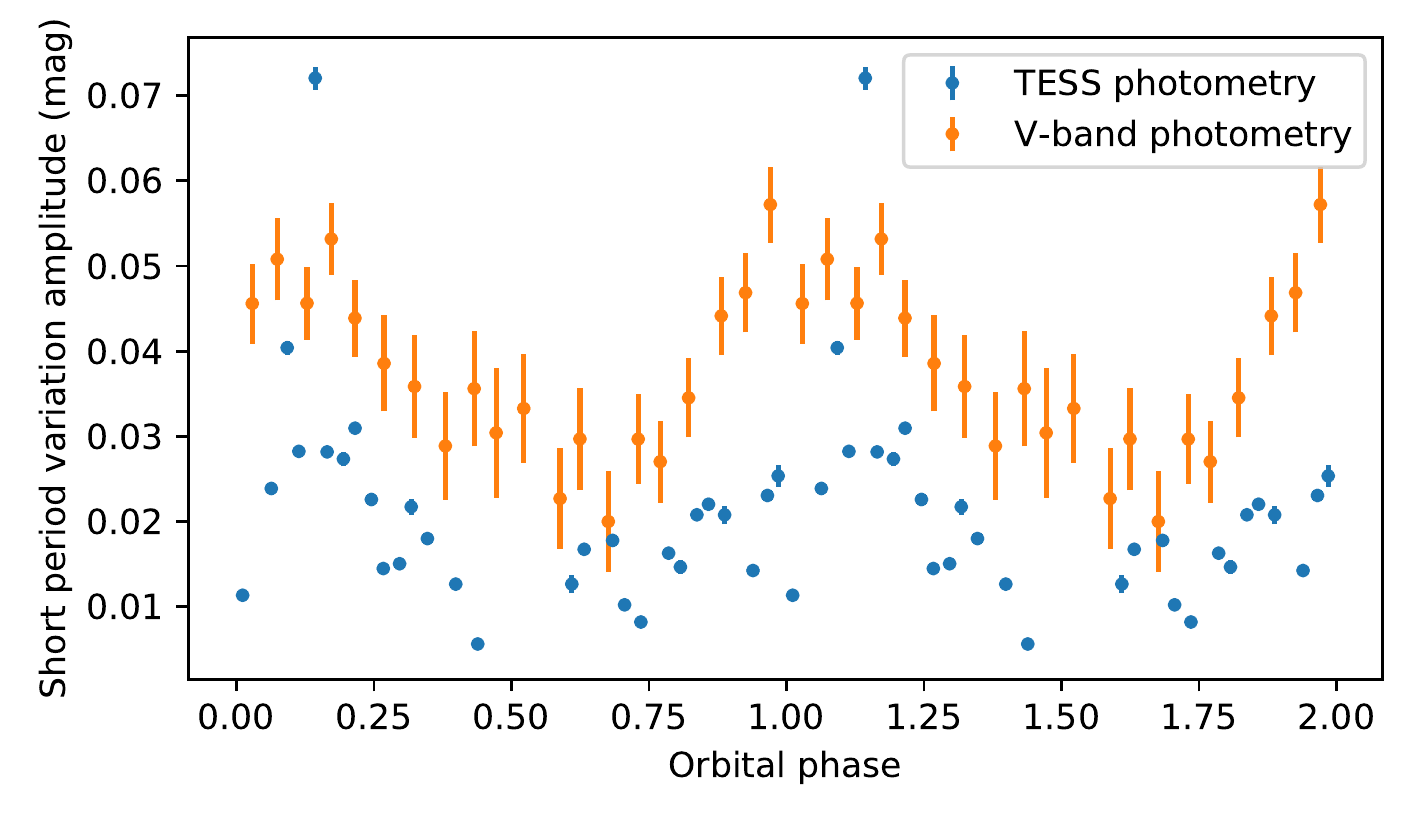}
    \caption{Amplitudes of the photometric 1.583-day period as a function of the orbital phase of the binary for both TESS and $V$-band data. Orbital phases are calculated using Equation \ref{eq:orbital_phase}. Figure \ref{fig:phot_short_amp_v_phase} demonstrates how the data for the $V$-band photometry variations were calculated.}
    \label{fig:tess_v_band_1583_amp_v_orbital}
\end{figure}

Figure \ref{fig:tess_v_band_1583_amp_v_orbital} clearly shows how the amplitude of the short 1.583-day variations is modulated throughout the orbital period of GG Car in the photometric data. For the TESS data a sinusoid is fitted to each 1.583-day slice within the dataset, with the amplitude and phase kept as free parameters. The amplitudes of the 1.583-day variations are then folded over the 31-day orbital period. It is clear that, in the TESS observational interval, the amplitude of the short-period variation is strongly tied to orbital phase, with the largest variations occurring around phase $\sim$0.1. This is shortly after the binary is at periastron, as per the ephemeris of Paper A, and when the system is brightest (both of which occur at phase 0). \\

For the $V$-band photometry, since there is not the fine time-sampling of the TESS data, different techniques need to be used to study the amplitude modulations of the short-period variations. To do this, the ASAS, ASAS-SN, and OMC data had the mean 31-day orbital variations subtracted, then a sliding window of orbital phase width 0.15 is used to calculate the short-period amplitudes in the orbital phase intervals. The data in each window were folded by the 1.583-day period and then fitted by a sinusoid, with amplitude, phase, and offset as free parameters. Figure \ref{fig:phot_short_amp_v_phase} in the Appendix demonstrates how the amplitude modulation of the $V$-band photometry along the binary orbit was calculated in each orbital phase window.\\

Figure \ref{fig:tess_v_band_1583_amp_v_orbital} shows that the amplitude of the short-period variations in the $V$-band data is also modulated across the orbital phase of the binary. The amplitude-phase signal closely matches that shown by the TESS data in phase and shape, demonstrating that this orbital dependence of the 1.583-day period is long-lived and persistent across all data, and is not a curiosity of the TESS observing epoch. It is also noteworthy that the sharp rise in the TESS amplitude is matched as the phase where the amplitude is largest in the $V$-band photometry, suggesting that the large spike in amplitude that is seen in the TESS data around phase 0.1 persists throughout the observations of GG Car. We find that the phases of the 1.583-day variations have no significant variability across the orbital period; this can be seen in the fitted sinusoids to the $V$-band data in Figure \ref{fig:phot_short_amp_v_phase}, which are all in phase. \\

In Figure \ref{fig:tess_orbital_subtracted}, it is clear that the variability in the TESS data is not uniform, but it does follow the general trend of having larger amplitudes shortly after periastron. The $V$-band data, since it is taking the average short-period amplitude of many orbital cycles simultaneously with a sliding window in orbital phase, gives us the average amplitude change over a long period of time. Therefore, we cannot conclude that this relationship holds for each orbital cycle, but that, in aggregate, over all orbital cycles there is a general trend for the amplitude of the photometric variability at the 1.583-period to be largest shortly after periastron. \\ 

\begin{figure} 
  \centering
    \includegraphics[width=.5\textwidth]{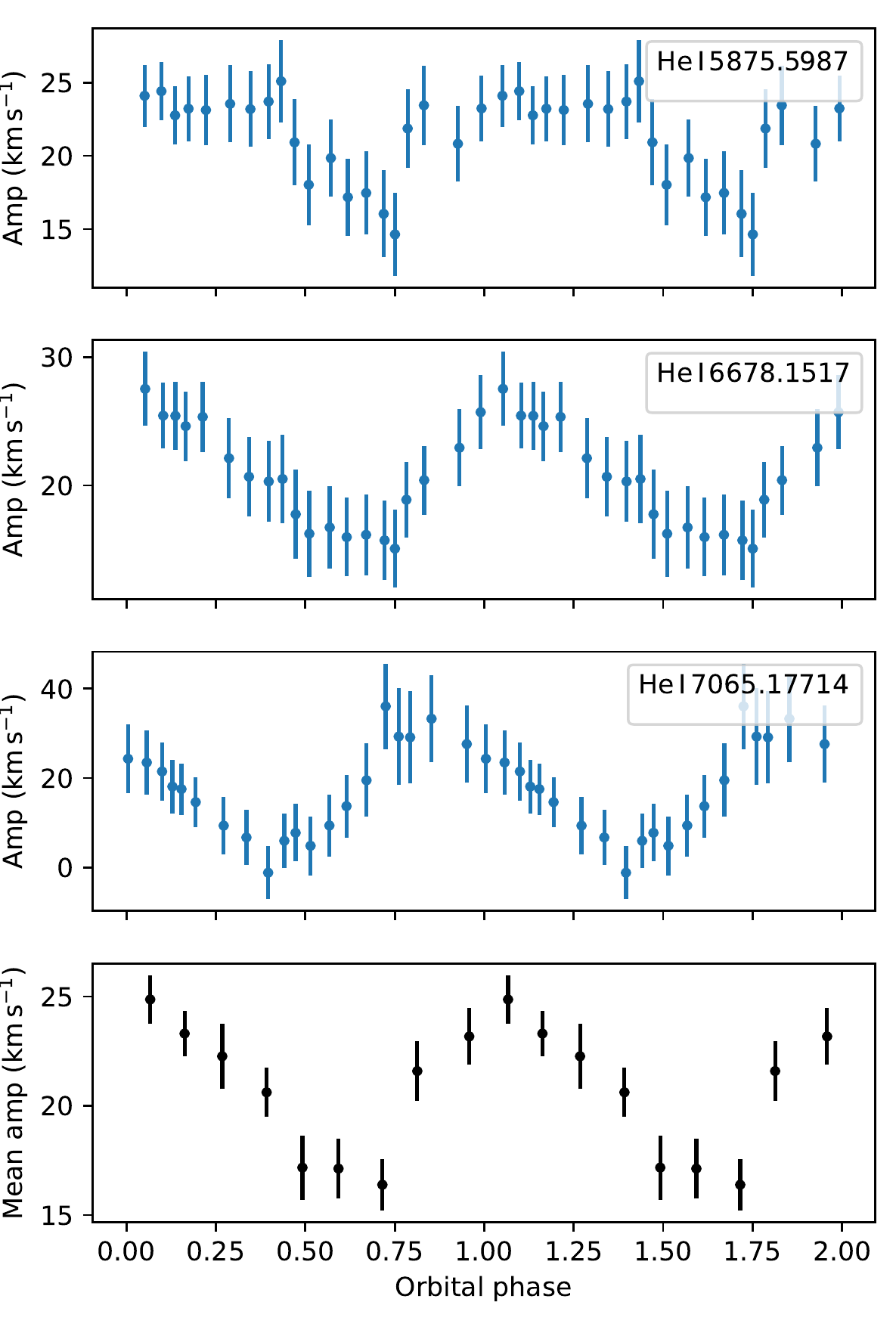}
    \caption{The amplitude of the short-period RV variations for the He\,I lines against orbital phase. The data were calculated in the same manner as the $V$-band data in Figure \ref{fig:tess_v_band_1583_amp_v_orbital}. The bottom panel shows the mean of the short-period amplitude in each phase bin, with the error-bar corresponding to the standard deviation in that bin.}
    \label{fig:he_amp_v_orbital}
\end{figure}

We now turn to see whether this amplitude modulation exists in the spectroscopic data. Figure \ref{fig:he_amp_v_orbital} displays the amplitude modulation along the binary orbit of the 1.583-day RV variations of the He\,I emission, as observed by the GJW. The amplitude modulations for the spectroscopy were calculated similarly to those of the $V$-band photometry. It is clear, for all three He\,I lines, that the amplitude of the spectroscopic RV variations are modulated in a similar manner to the photometric variations.  He\,I 7065, however, has an offset to the other He\,I lines and the photometry with peak amplitude occurring around phase 0.9.\\

\begin{figure} 
  \centering
    \includegraphics[width=.5\textwidth]{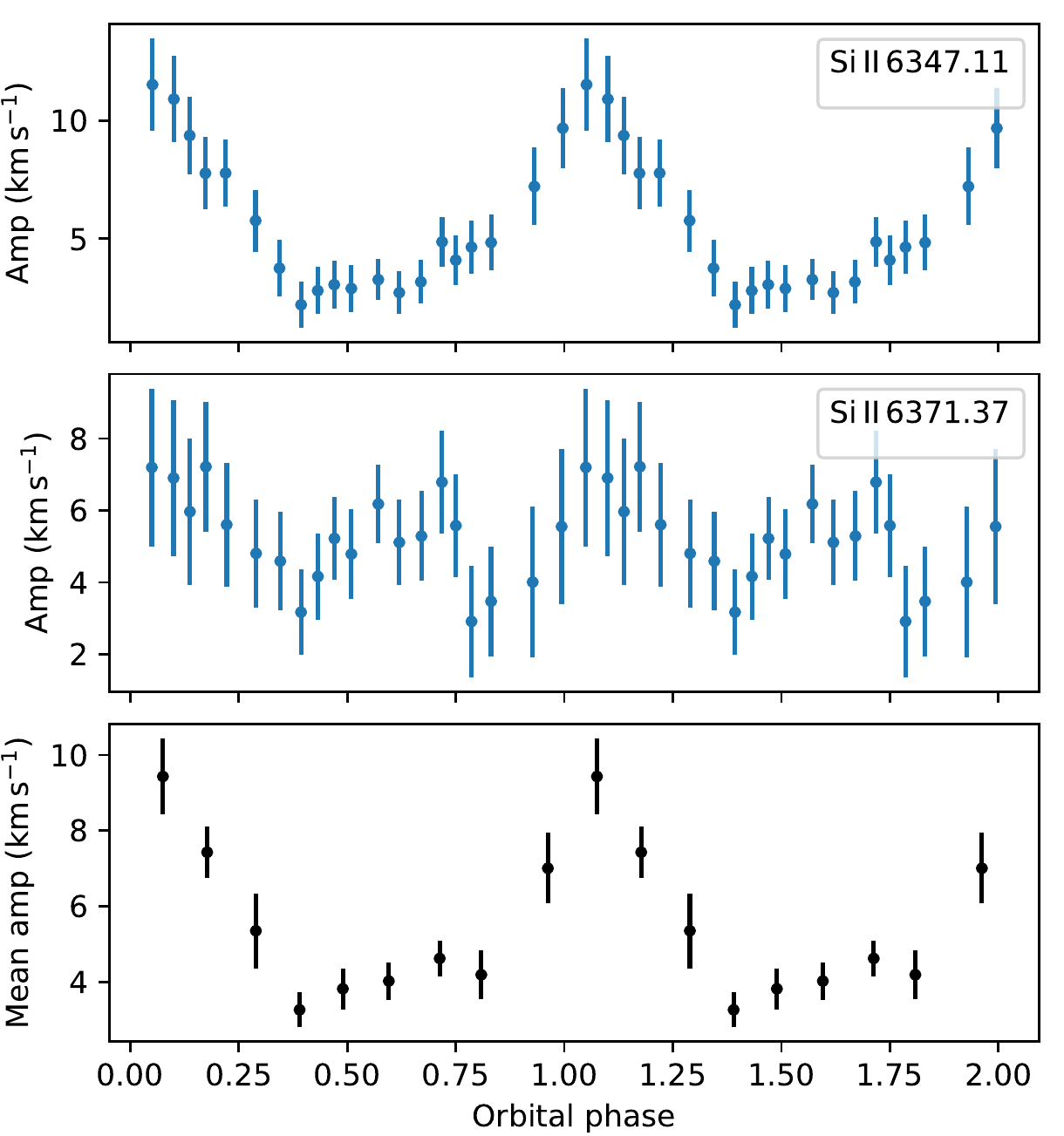}
    \caption{Same as Figure \ref{fig:he_amp_v_orbital}, except for the Si\,II 6347 and 6371 lines.}
    \label{fig:si_amp_v_orbital}
\end{figure}

\begin{figure} 
  \centering
    \includegraphics[width=.5\textwidth]{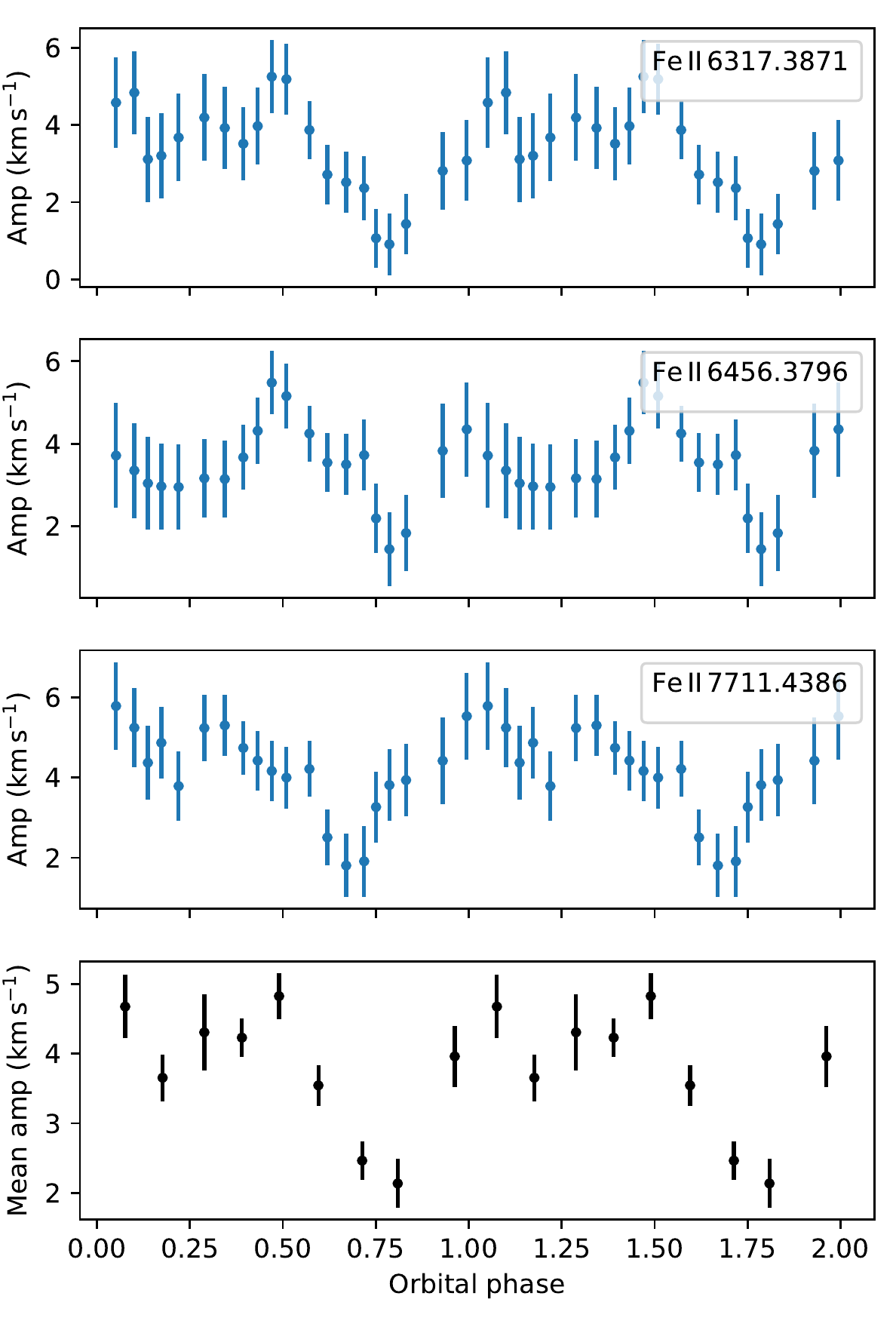}
    \caption{Same as Figure \ref{fig:he_amp_v_orbital}, except for the Fe\,II lines which display the short period variability.}
    \label{fig:fe_amp_v_orbital}
\end{figure}

Figures \ref{fig:si_amp_v_orbital} and \ref{fig:fe_amp_v_orbital} show the same for the short-period presenting Si\,II and Fe\,II lines respectively. The Si\,II 6347 line shows a clear signal, with the amplitude of the short-period variations modulating across the orbital phase in the same manner as the photometry and the He\,I, whereas Si\,II 6371's signal of amplitude modulation is less clear. The Fe\,II shows some indication of amplitude modulation over the binary orbit, however their signals are not very clear due to the smaller amplitudes of the Fe\,II lines. The amount of amplitude modulation is largest for the He\,I lines, followed by the Si\,II lines, followed by the Fe\,II lines. \\

\section{Discussion}
\label{sec:discussion}

\subsection{Origin of the 1.583-day variability}
\label{sec:1583_origin}
Stellar variability is variously explained by rotations, multiplicity, or pulsations. As evidenced by the TESS data, the amplitude of the photometric variability at 1.583 days can reach up to 0.07\,mag. This implies a peak-to-trough brightness change of $\sim$14\,\%; since the secondary of GG Car is predicted to contribute <\,3\% of the flux of the system assuming it is a main sequence star of 7.2\,$M_\odot$ (Paper A), then the flux from the secondary alone cannot be the origin of this variability. Here, we investigate whether rotation of the primary, a hidden third body, or pulsations in the primary can cause the 1.583-day variability observed in GG Car.\\ 

As shown in \cite{Zickgraf1996SpectroscopicClouds} the critical rotation velocity at the equator of a star, above which it cannot rotate without breaking up, can be estimated by
\begin{equation}
    v_{\rm crit} = \sqrt{\frac{G(1-\Gamma_{\rm rad})M}{R}},
    \label{eq:v_crit}
\end{equation}
\noindent where $G$ is the gravitational constant, $M$ is the mass of the star, $R$ is the radius of the star, and $\Gamma_{\rm rad}$ is the correction to the effective gravity due to radiation pressure by electron scattering. $\Gamma_{\rm rad}$ is given by
\begin{equation}
    \Gamma _{\rm rad} = \frac{\sigma_e L}{4\pi G M c},
    \label{eq:Gamma_rad}
\end{equation}
\noindent where $L$ is the stellar luminosity, $c$ is the speed of light, and $\sigma_e$ is the electron mass scattering coefficient. For $\sigma_e$, we adopt a value of 0.308\,$\text{cm}^2 \, \text{g}^{-1}$ taken from \cite{Lamers1986TheStars}; it should be noted that this value of $\sigma_e$ is calculated for the composition of the circumstellar environment of P Cygni. Entering the stellar parameters of GG Car given in Table \ref{tab:ggcar_parameters} gives $v_{\rm crit} = 370\pm70$\,\kms. Should the 1.583-day period be interpreted as the rotation period of the \sgBeshorthand \ primary, the surface rotation velocity would be $v_{\rm rot} = 860\pm260$\,\kms at the equator, far exceeding the critical rotation velocity of the star. Clearly the 1.583-day period cannot be the stellar rotation period of the \sgBeshorthand\ primary component of GG Car. \\

Might there be a hidden close companion of the \sgBeshorthand\ primary, or circumstellar material which orbits the primary every 1.583 days, causing the short-period variability of the system? Kepler's third law states

\begin{equation}
    P^2 = \frac{4\pi^2a^3}{GM},
\end{equation}

\noindent where $P$ is the orbital period, $a$ is the semi-major axis of the orbit, and $M$ is the total mass of this inner system. Entering $P = 1.58315\pm0.0002$\,days, and the primary's mass of $M = 24\pm4\, M_\odot$ gives $a=16.5\pm0.9\,R_\odot$. Given that GG Car's primary has a radius of 27\,$R_\odot$, this implies that a hidden companion would be completely engulfed in the primary star. Whilst it is theorised that \sgBeshorthand s may be post-merger objects \citep{Podsiadlowski2006corrected}, the variable period would not be as stable as it is observed to be if GG Car were a recently-merged object, given that an indication of a $\sim$1.6-day period was first reported by \cite{Gosset1984} and the periodicity continues to the present day.\\

This leaves pulsations as the likely cause of the short-period variability in GG Car. Pulsations have been observed in B[e] stars at a similar timescale to the one we observe in GG Car, though they are rare. \cite{Krtickova2018AnEnvelopes} detects a pulsation period of $1.194 \pm 0.06$\,days in the unclassified B[e] star HD\,50138 (V743 Mon). Pulsations with periods at this timescale are often observed in blue supergiants (e.g. \citealt{Haucke2018WindSupergiants}). \cite{Saio2013EvolutionAbundances} reports that radial pulsations and a spectrum of non-radial pulsations may be excited in evolved blue supergiants (BSGs) which have already undertaken the blue loop in their post-main sequence evolution, having been in a prior red supergiant (RSG) state. Conversely, they find that most of these pulsations were suppressed in BSGs which had not yet undergone the blue loop. Since we only observe one significant frequency, other than the orbital frequency, in the periodograms of the photometry and spectroscopy of GG Car this likely indicates that the primary GG Car is in a pre-RSG state, according to the conclusion of \cite{Saio2013EvolutionAbundances}. This supports the findings of \cite{Kraus2009} and \cite{Kraus2013}, which concluded that GG Car is in a pre-RSG state based on $^{13}$CO abundances. \\

\textcolor{black}{Given that the variability is coupled to the orbit of the binary, this may indicate that the tidal potential is exciting pulsations in the \sgBeshorthand\ primary (this amplitude modulation is discussed in further detail in Section \ref{sec:amplitude_modulation}). As the tidal potential is quadrupolar, the most likely oscillation mode we are observing is an l=2 mode, where $l$ is the degree of the mode and indicates the number of surface nodes. \cite{Gough1993LinearPulsation} shows that f-modes of stars, which act as surface gravity waves with $n = 0$ (where $n$ is the number of radial nodes of the oscillation mode), have angular frequencies which may be determined as} \\

\begin{equation}
\label{eq:l2fmode_frequency}
    \omega^2 = \frac{L g_s}{R}\left(1 - \epsilon(L)\right),
\end{equation}

\noindent \textcolor{black}{where $\omega = 2\pi / P$ is the angular frequency of the mode, $g_s = GM/R^2$ is the surface gravity of the star, $R$ is the stellar radius, $L = \sqrt{l (l + 1)}$, and $\epsilon(L)$ is a term to correct for the sphericity of the star. $\epsilon(L)$ is calculated}

\begin{equation}
\label{eq:l2fmode_frequency_correction}
    \epsilon (L) = 2L^{-1} + \frac{3\int^{R}_{0}(r/R - 1) \rho \exp\left(2Lr/R\right)\,{\rm d}r}{\int^{R}_{0}\rho \exp\left(2Lr/R\right)\,{\rm d}r},
\end{equation}

\noindent \textcolor{black}{where $\rho = \rho(r)$ is the density of the star. Entering the stellar parameters of GG Car into Equation \ref{eq:l2fmode_frequency}, and utilising simple distributions of $\rho(r)$, yields values of the $l=2$ f-mode frequency which are consistent with our observations. Assuming a constant density gives $P = 2.4^{+1.2}_{-1.0}$\,days. While a constant density is highly unrealistic, we may use simple prescriptions to give higher densities in the stellar centre, such as $\rho(r) \propto 1 - A\,(r/R)^{B}$, where $A$ and $B$ are constants, that also yield consistent periods. Computing a grid of allowed periods for the $l=2$ mode using this simple prescription of $\rho(r)$ returns periods which are consistent with the observed periodicity for all values of $A$ and $B$ where $0 \leq A \leq 1$ and $B > 0$. While detailed modelling is beyond the scope of this paper, this shows that the observed pulsation frequency is consistent with and likely to be the $l=2$ f-mode. It must also be noted that, per Equation \ref{eq:l2fmode_frequency}, higher values of $l$ up to $\sim$8 may also yield periods which are consistent with the observed variability; however, the $l=2$ mode would be expected to be excited more strongly than the higher modes by the tidal potential. The mode observed may not be radial, since tidal modulation is only allowed for pulsation modes with $l \neq 0$ \citep{Polfliet1990DynamicOscillations.}.}\\

The RV variability that we detect in GG Car's emission lines would then be related to that pulsations in the primary affecting the structure of the wind at its 1.583-day periodicity. Pulsations have been theorised and shown to affect the wind and mass-loss of properties of blue supergiant stars \citep{Aerts2010Periodic50064, Kraus2015Interplay478, Yadav2016Stability198478, Yadav2017InstabilityStars, Haucke2018WindSupergiants}. Structures in stellar winds caused by pulsations have been proposed to explain variability in certain X-ray binaries \citep{Finley1992Periodic0114+65., Koenigsberger2006The010}. \\

\subsection{1.583-day amplitude modulation}
\label{sec:amplitude_modulation}

In Section \ref{sec:orbital_phase_dependence_of_short_variability}, we have shown that the amplitude of the 1.583-day variations is modulated by the orbital phase of the binary, most clearly in the photometry. The amplitudes are largest when the binary is at periastron in its eccentric ($e=0.5\pm0.03$) orbit. This linking between the 31-day orbital period and the 1.583-day short period is unusual, since the ratio between the two periods is $31.01 / 1.583 = 19.589$, i.e. they are non-commensurate. \\

GG Car's lightcurve bears a resemblance to the ``heartbeat stars'' which have resonant, tidally driven, stellar oscillations that have variable amplitudes over the orbital period (see e.g. \citealt{Fuller2017HeartbeatLocking}). Most heartbeat stars yet discovered are lower mass A and F stars, but the phenomenon has also been observed in massive O and B stars \citep{Pablo2017TheOrionis, Jayasinghe2019AnTESS}. These heartbeat stars are eccentric binaries which have orbital periods that are exact integer multiples of the star's g-mode pulsation periods, which lead to coherent and resonant pulsations due to the tidal excitation of the oscillation modes (see also \citealt{Kumar1995TidalPulsars, DeCat2000AFrequencies, Willems2002TidallyBinaries}). However, the short-period variability we observe in GG Car is clearly non-resonant with the orbital period, as the bottom panel of Figure \ref{fig:tess_orbital_subtracted} and the non-integer relation between the short-period and orbital period clearly show. Therefore, the periodicity and amplitude modulation we report cannot arise due to resonance. However, there are indications that tidal effects can affect non-resonant free oscillations, and here we explore that possibility.\\

Paper A showed that, at periastron, the radius of the primary of GG Car extends to $85 \pm 28$\,\% of its Roche lobe radius, whereas at apastron it only extends to $28 \pm 9$\,\%; therefore, the tidal perturbation at periastron will be significant and a dynamical tide will be raised. Paper A also presents evidence that the primary's mass-loss is focused around periastron. In GG Car, the timescale of the varying gravitational potential will be short given that the orbit is significantly eccentric, with the timescale of periastron passage $T_{\rm peri} \sim \sqrt{a^3(1-e)^3 / G M_{\rm tot}} \sim 1.7$\,days, where $M_{\rm tot}$ is the combined mass of the primary and the secondary. A similar determination of the timescale of intense gravitational interaction is the half-width-at-half-maximum (HWHM) of the tidal force, i.e. the time taken for the tidal force of the secondary on the primary to increase from mid-point value to its peak value at periastron. The HWHM of the tidal force is $\sim$1.9\,days (since $F_{\rm tidal} \propto r^{-3}$, where $r$ is the instantaneous separation of the binary components). Therefore, around periastron, the timescale of the change of the tidal force is of the same order as the observed periodicity and dynamical timescale of the primary; the star will be unable to adjust to the changing tidal force in a quasi-static way. Conversely, for the tidal force to go from its minimum at apastron to the mid-point value takes 13.6\,days, i.e. an order of magnitude longer than the pulsation period and the dynamical timescale. \\

It therefore follows that the varying proximity of the two binary components will affect the conditions of the primary, and the rapid change of the tidal forces and enhanced mass loss at periastron will draw the primary out of hydrostatic equilibrium at a timescale of the same order as its dynamical timescale. The primary, attempting to regain equilibrium, oscillates at the observed period of 1.583 days, \textcolor{black}{which we have shown in Section \ref{sec:1583_origin} is likely to be the $l=2$ f-mode}. As the binary components separate after periastron passage, the tidal force becomes increasingly less important, and the star will continue to oscillate at the observed period and ultimately try to regain hydrostatic equilibrium. The timescale for the damping of the oscillations depends on the dominant source of viscosity, \textcolor{black}{but, in the case of the large tidally induced distortion observed in GG Car, could be as fast as the dynamical timescale of the primary's envelope ($\tau_{\rm dyn} \sim 1$\,day). Quantifying mode damping timescales in the envelopes of massive OB stars is an uncertain problem, and is beyond the scope of this paper.} \\

\begin{figure} 
    \centering
    \includegraphics[width=0.5\textwidth]{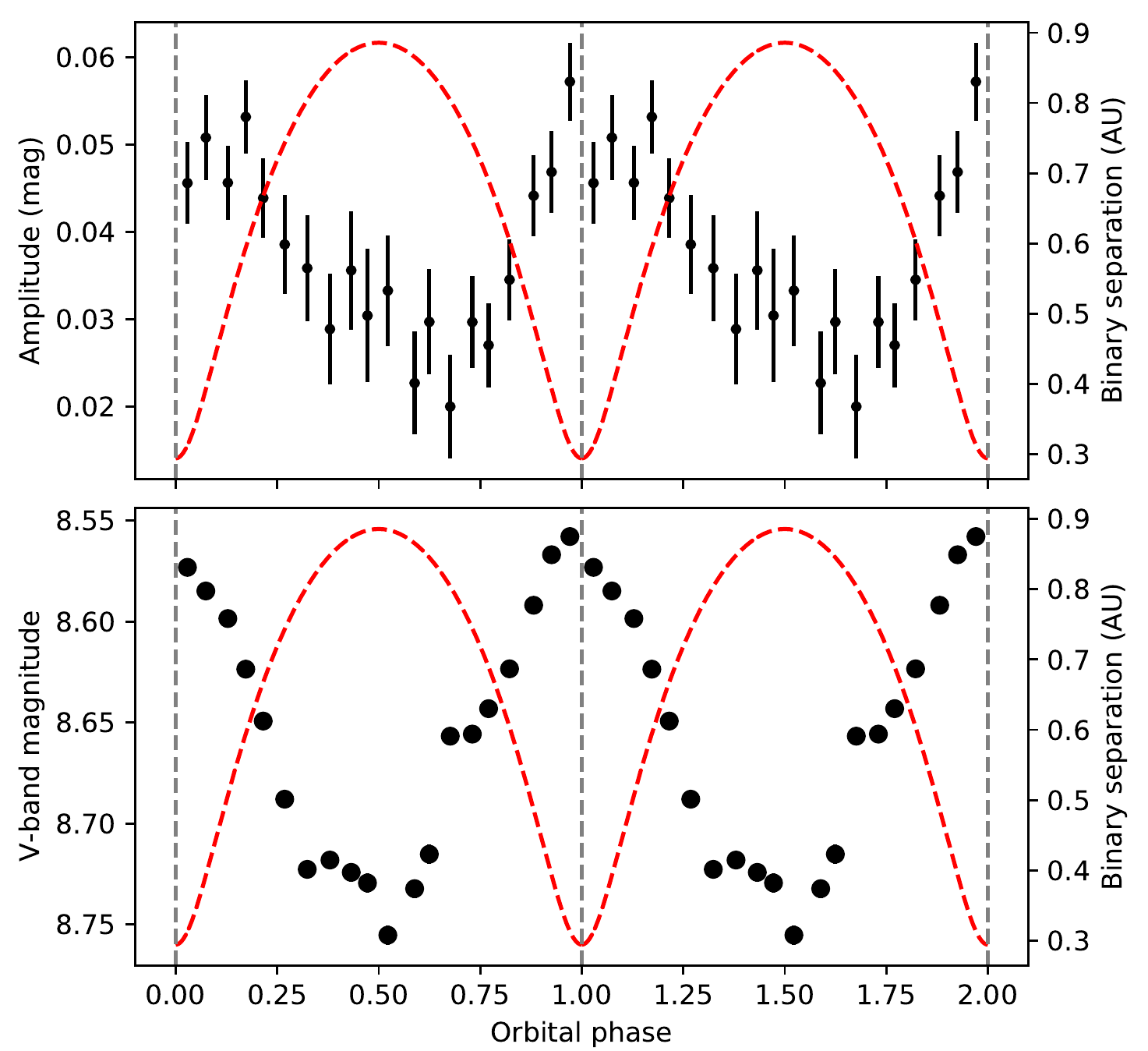}
    \caption{Top: Amplitude of the 1.583-day variations in $V$-band photometry against orbital phase (black points, left hand axis), with the instantaneous separation of the binary components over-plotted (red dashed line, right hand axis). Phases of periastron are denoted by vertical dashed lines.
    Bottom: same as top, except with the mean $V$-band magnitude in the phase bin replacing the short-period amplitudes.}
    \label{fig:amplitude_flux_separation}
\end{figure}

Figure \ref{fig:amplitude_flux_separation} displays both how the short-period variation amplitude and mean brightness in the $V$-band compares with the instantaneous separation of the binary components along the orbital period: the mean brightness is near perfectly anti-correlated with the separation of the components. On the other hand, the short-period amplitude of the $V$-band data increases more rapidly than it decays, and peaks somewhere between orbital phases 0.0 and 0.14, corresponding to $\sim$0\,--\,4\,days after periastron. This delay can also be clearly seen in the folded TESS data in Figure \ref{fig:tess_orbital_subtracted}. \\ 

There are examples in the literature which support this hypothesis of tidally modulated free oscillations. \cite{Moreno2011EccentricBinaries} calculated that increased stellar activity can be expected on stellar surfaces around periastron in eccentric binaries due to the raising of dynamical tides and the associated changes in timescale of dissipation of the tidal energy, and this can lead to oscillations. The interaction of free oscillations with tidal interaction was theoretically studied by \cite{Polfliet1990DynamicOscillations.}, who show that a tidally distorted star may display free non-radial oscillations with periods of the order of its dynamical timescale. They find that the free oscillations' amplitudes are modulated at a frequency which is an integer multiple of the orbital frequency. Tidal modulation of pulsation amplitudes in this manner have been reported in the $\beta$ Cephei variables $\beta$ Cep \citep{Fitch1969EvidenceCephei}, CC And \citep{Fitch1967EvidenceScorpii}, $\sigma$ Scorpii \citep{Fitch1967EvidenceScorpii, Goossens1984LightScorpii., Chapellier1992PulsationScorpii.}, $\alpha$ Vir \citep{Dukes1974TheSpica.}, and 16\,Lacertae \citep{Fitch1969EvidenceCephei,Chapellier1995PulsationLacertae.}; in these objects, the pulsation periods and the orbital periods are non-resonant, but the pulsational amplitudes undergo an integer number of cycles per orbital period. \cite{Chapellier1995PulsationLacertae.} reports that the amplitude of an $l=1$ pulsation mode of the system 16\,Lacertae undergoes exactly one cycle over the orbital period where the pulsation period and the orbital period are non-commensurate, similar to what we observe in GG Car.\\

\textcolor{black}{It is worth noting that the amplitude modulation that we are reporting in GG Car also bears resemblance to binaries which have been recently discovered to have pulsations that are tidally trapped on one hemisphere of the variable component. \cite{Handler2020TidallyTESS} discovered a tidally-trapped pulsation mode in the binary star HD\,74423, in the form of amplitude modulation of the observed pulsations as a function of orbital phase. Similarly to what we observe in GG Car, the pulsation frequency and orbital frequency in HD\,74423 are non-commensurate. \cite{Kurtz2020TheCamelopardalis} find a similar result in CO Cam, finding that four modes are trapped by the tidal potential of the companion. \cite{Fuller2020TidallyStars} presents evidence of a similar process occurring in TIC\,63328020. The authors explain the amplitude modulation in these systems as being due to the pulsation axis of the variable component to be aligned with the line of apsis of the binary which cause the pulsations to have a larger amplitude on one hemisphere of the star, either the hemisphere facing towards or away from the companion. A larger, or smaller, photometric variability amplitude is then observed at times of conjunction depending on which hemisphere is facing the observer. According to the orbital geometry of GG Car ($e = 0.50$, $\omega = 339.87^\circ$), superior conjunction occurs at phase 0.14 and inferior conjunction occurs at 0.93. Superior conjunction, therefore, does occur at a remarkably similar phase as the oscillation amplitude's maximum, most clearly shown in Figure \ref{fig:tess_orbital_subtracted}. However, in the scenario of tidally trapped pulsations, the phase of inferior conjunction would then be expected to have the lowest oscillation amplitude. This is clearly not the case, as the oscillation amplitudes are still very large around phase 0.93. Therefore it is unlikely that the amplitude modulation we observe in GG Car is due to tidal trapping of the pulsation mode, though geometrical effects may perhaps be accentuating the observed amplitude of the non-radial pulsation mode at superior conjunction.}\\

Further TESS-quality observations of GG Car observing more orbital periods would be needed to fully confirm such an argument of orbital-phase modulated free oscillations. Alternatively, phase-resolved studies of the spectral energy distribution (SED) of the system could allow the varying contribution of the primary to the SED to be measured. Should the primary's contribution vary with the orbit and the 1.583-day period, this would lock down the variability as being due to pulsations of the primary, and therefore the amplitude modulation would be due to proximity effects of the primary to the secondary.\\

\section{Conclusions}
\label{sec:conclusions}
We have shown that the \sgBeshorthand\ binary GG Car is significantly variable in both photometry and spectroscopy at $1.583156 \pm 0.0002$\,days, and we have studied this variability in detail for the first time. This period is much shorter than the well-known 31-day orbital period of the binary. We have shown that the short-period variability cannot be caused by the rotation of the \sgBeshorthand\ primary, the presence of a hidden third body, or intrinsic variability of the secondary's flux. \textcolor{black}{We find that 1.583\,days is consistent with the period of the lower-order f-modes ($l \lessthanapprox 8$) of GG Car's primary given its mass and radius, and we ascribe the variability as most likely being due to the $l=2$ f-mode such that it couples to the quadrupolar tidal potential. We therefore argue that pulsations of this mode are the most likely cause of its variability.} \\

In spectroscopy, we found that the short period manifests itself in the RVs of the He\,I, Si\,II and Fe\,II emission lines; however, not all of GG Car's emission lines display the periodicity. We have found that the amplitudes of the spectroscopic RV variations at the 1.583-day period are correlated with the upper energy levels of the transitions causing the line emission, implying that the variations are related to the temperature of the line forming regions. \\

We have shown that the amplitudes of the short-period variations are dependent on the orbital phase of the binary, most notably for the $V$-band and TESS photometry, with the largest variations occurring around or just after periastron, where the system is also at its brightest. This is striking as the ratio between the orbital period and the shorter period is 19.596. This non-integer ratio of the two periods means the shorter period cannot be a tidally-resonant excited oscillation mode in one of the stars. \\

Paper A shows that the primary's radius extends to $\sim$85\% of its Roche radius at periastron, compared to only $\sim$28\% at apastron. We have shown, in Section \ref{sec:amplitude_modulation}, that the timescale of the change of the tidal forces on the primary at periastron are of the same order as its dynamical timescale. \textcolor{black}{Therefore, we suggest that the primary is being pulled out of hydrostatic equilibrium by the secondary every orbit due to the strong tidal effects at periastron faster than the primary can regain equilibrium. This loss of equilibrium causes pulsations at the $l=2$ f-mode, which can couple to the quarupolar tidal potential and which is consistent with the 1.583-day period observed, with a larger amplitude when the stars are close in proximity and the primary is being pulled further from equilibrium.} These oscillations are damped \textcolor{black}{at a timescale which may be as fast as the dynamical timescale} as the separation between the binary components increases and the primary can return to equilibrium. \\

The unusual behaviour of GG Car's short-period variability reported in this paper has not been reported in other \sgBeshorthand s as of yet. Further TESS-quality observations and phase-resolved SED observations of GG Car would be required to pin down the cause of its short-period variability and amplitude modulation, and similar phenomena in other \sgBeshorthand s in binaries should be searched for.\\

\section*{Acknowledgements}
We thank John Papaloizou for his useful discussions. AJDP thanks the Science \& Technology Facilities Council (STFC) for their support in the form of a DPhil scholarship. Part of this work was based on data from the OMC Archive at CAB (INTA-CSIC), pre-processed by ISDC. A great many organisations and individuals have contributed to the success of the Global Jet Watch observatories and these are listed on {\tt www.GlobalJetWatch.net} but we particularly thank the University of Oxford and the Australian Astronomical Observatory. This research has made use of NASA's Astrophysics Data System. This research has made use of the SIMBAD database, operated at CDS, Strasbourg, France.  This work made use of data supplied by the UK Swift Science Data Centre at the University of Leicester.

\section*{Data availability}
ASAS $V$-band photometric data available from \url{http://www.astrouw.edu.pl/cgi-asas/asas_cgi_get_data?105559-6023.5,asas3}.\\
ASAS-SN $V$-band photometric data available from \url{https://asas-sn.osu.edu/}.\\
OMC $V$-band photometric data available from \url{https://sdc.cab.inta-csic.es/omc/secure/form_busqueda.jsp}.\\
TESS FFI data was accessed and reduced via the \texttt{eleanor} framework \citep{Feinstein2019Eleanor:Images}; the \texttt{python3.x} reduction code used to access the data presented in this article will be shared on reasonable request to the corresponding author.\\
The fits to spectroscopic Global Jet Watch data underlying this article will be shared on reasonable request to the corresponding author.

\bibliography{references_new}{}
\bibliographystyle{mnras}

\appendix

\section{Amplitudes of spectroscopic radial velocity variations}
\label{sec:appendix_k}
To determine the RV variability of the emission lines, Keplerian orbital RV solutions are fitted to the RV data for each line separately at both the orbital and 1.583-day periods simultaneously, fitting for the amplitude $K$, the eccentricity $e$, the argument of periapsis $\omega$, and the phase of periastron $M_0$ for each period. We also fit for jitter, $j$, modelled as a correction of the RV uncertainties. We also fit the systemic velocity $v_0$ separately for each line. The fits at the orbital period are discussed in Paper A, and are used to determine the orbital solution of the binary. We show in Section \ref{sec:discussion} that the 1.583-day period cannot be an orbital period of a hidden inner binary; however, modelling the RV variations with a Keplerian solution is useful for fitting an amplitude to a repeating signal of an arbitrary shape in noisy data. We can then mutually compare the amplitudes found between the line species.\\

\begin{table} 
\centering          
\begin{tabular}{ l l}
\hline \hline
Line	&	$K$	(\kms) \\ \\
Fe$\,$II$\,$6317.3871	&	$3.26^{+0.88}_{-0.86}$	\\
Fe$\,$II$\,$6456.3796	&	$3.49^{+0.97}_{-0.7}$	\\
Fe$\,$II$\,$7711.4386	&	$4.06^{+0.75}_{-0.78}$	\\
He$\,$I$\,$5875.5987	&	$20.8^{+2}_{-1.8}$	\\
He$\,$I$\,$6678.1517	&	$20.1^{+2.2}_{-1.9}$	\\
He$\,$I$\,$7065.17714	&	$14.9^{+7.3}_{-5}$	\\
Si$\,$II$\,$6347.11	&	$5.51^{+1}_{-0.9}$		\\
Si$\,$II$\,$6371.37	&	$5.57^{+1.8}_{-1.3}$	\\

\end{tabular}
\caption{Amplitudes, $K$, of the 1.583-day RV variations for each emission line which displays variations at this period.} 
\label{tab:orbital_solutions_short}      
\end{table}

We fit the RV variations for each emission line separately by maximising the log-likelihood function, using the Monte Carlo Markov Chain algorithm \texttt{emcee} \citep{Foreman-Mackey2012Emcee:Hammer}. The log-likelihood for a set of parameters, $\theta$, given $N$ RV data points, $D$, with uncertainty $\sigma$ is given by
\begin{multline}
\label{eq:log_like}
\begin{split}
    \text{ln} \, P (\theta \mid D, \, \sigma) = - \frac{1}{2} \sum_{i=0}^N \left[\frac{\left(D_i - v_{\rm kep}(\theta_{\rm orb}) - v_{\rm kep}(\theta_{1.583}) - v_0\right)^2}{\sigma_i^2 + j^2}  \right.\\ \left.+ \text{ln}\left(2\pi (\sigma_i^2 + j^2)\right)\right],
\end{split}
\end{multline}
\noindent where $\theta_{\rm orb}$ and $\theta_{1.583}$ are the orbital parameters for the long- and the short- period respectively, and $v_{\rm kep}$ is the Keplerian velocity calculated for a set of orbital parameters. The fitted parameters, $\theta$, encode $\theta_{\rm orb}$, $\theta_{1.583}$, $v_0$, and $j$. The uncertainties, $\sigma$, are taken from the least-squares fitting algorithm of the Gaussian fitting routines. Table \ref{tab:orbital_solutions_short} lists the fitted amplitudes for all emission lines in this study at the 1.583-day period. The fitted parameters at the orbital period are listed in Paper A, appendix B.\\

Although we are not implying that the 1.583-day period is in any way due to an orbital effect, fitting Keplerian RV solutions is a convenient method to fit an arbitrarily shaped, periodic signal in this context, which allows for effective extraction of mutually comparable amplitudes. We ignore all parameters encoded in $\theta_{1.583}$ other than $K$, as they are fitted for convenience only and will have no physical significance.\\

\section{Calculation of phase-amplitude figures}
\begin{figure*} 
  \centering
    \includegraphics[width=\textwidth]{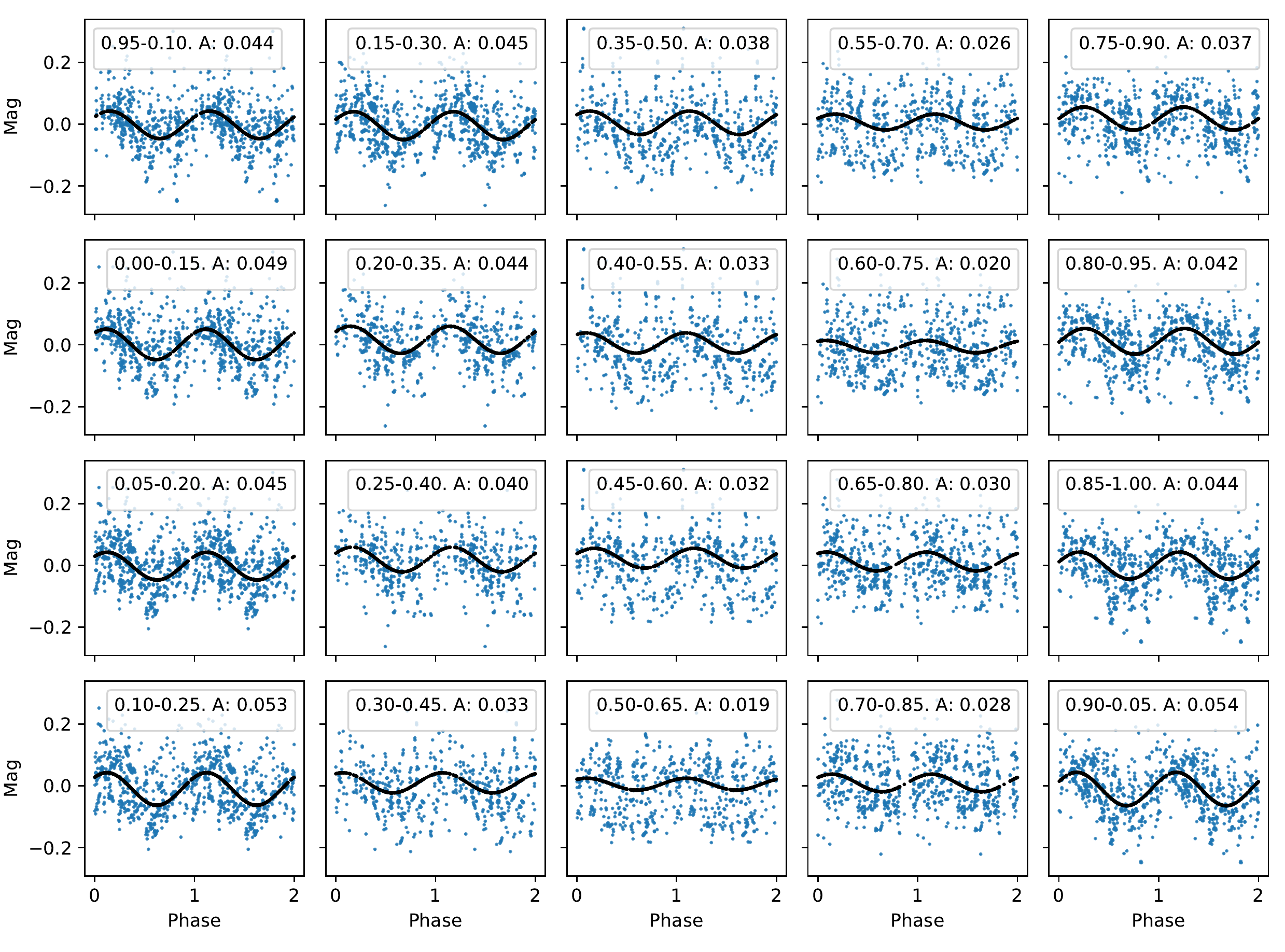}
    \caption{Demonstration of how the short-period amplitude versus orbital phase for the $V$-band photometric data, shown in Figure \ref{fig:tess_v_band_1583_amp_v_orbital}, was calculated. Each panel shows the $V$-band data, with the 31-day orbital variance subtracted, in an orbital phase window of width 0.15 folded by the 1.583-day period. A black line shows the sinusoid fitted to the data in that phase window. The legend in each panel says the start phase, the end phase, and the amplitude of the fitted sinuosoid. The phase of the windows increases first down the columns in the figure, and then along the rows.}
    \label{fig:phot_short_amp_v_phase}
\end{figure*}

Figure \ref{fig:phot_short_amp_v_phase} demonstrates how the short-period amplitude versus orbital phase for the $V$-band photometric data, shown in Figure \ref{fig:tess_v_band_1583_amp_v_orbital}, was calculated. The $V$-band data had the photometric variations at the orbital period subtracted, then were binned by a sliding window in orbital phase, of phase width 0.25. The data within a window are then folded by the 1.583-day short period, and a sinusoid is fitted and the amplitude extracted. Each panel shows the photometric data in blue and the sinusoid fits as black in each of the phase windows. \\

The amplitudes of the 1.583-day variations vary significantly with orbital phase. The phase of the sinusoidal variations do not change significantly given the uncertainties of the fitted parameters.\\

\bsp	
\label{lastpage}
\end{document}